%% file: arXiv.tex
\documentclass[traditabstract]{aa} %

\newcommand{\Ni}{\ensuremath{^{56}\mathrm{Ni}}}
\newcommand{\Co}{\ensuremath{^{56}\mathrm{Co}}}

\newcommand{\Msun}{\ensuremath{\mathrm{M}_\odot}}
\newcommand{\Msunpyr}{\ensuremath{\Msun~\mathrm{yr^{-1}}}}
\newcommand{\kmps}{\ensuremath{\mathrm{km~s^{-1}}}}

\def\ifundefined#1{\expandafter\ifx\csname#1\endcsname\relax}
\ifundefined{ensuremath}\def\ensuremath#1{\relax\ifmmode{#1}}
\else${#1}$\fi\else\relax\fi
\ifundefined{nuc}\def\nuc#1#2{\relax\ifmmode{}^{#1}{\protect\mathrm{#2}}
\else${}^{#1}$#2\fi}\else\relax\fi
\usepackage{txfonts}

\usepackage{natbib}
\usepackage{deluxetable}
\usepackage{url}
\usepackage{graphicx}

\bibpunct{(}{)}{;}{a}{}{,} 

\usepackage[normalem]{ulem}
\makeatletter
\renewcommand{\fnum@figure}{Fig. \thefigure.}
\makeatother

\graphicspath{{./}{figures/}}

\begin{document} 

   \title{The Carnegie Supernova Project II. \\ Observations of SN~2014ab possibly revealing a 2010jl-like SN~IIn with pre-existing dust\thanks{This paper includes data obtained  with  the 6.5-m Magellan Telescopes located at Las Campanas Observatory, Chile, and the ESO Paranal Very Large Telescope under ESO program  092.D-0645.}$^{,}$\thanks{Photometry and spectra presented in this paper are available on WISeREp.}}

\author{T. J. Moriya\inst{1,2}
\and
 M. D. Stritzinger\inst{3}
\and 
F. Taddia\inst{3}
\and 
N. Morrell\inst{4}
\and 
N. B. Suntzeff\inst{5,6}
\and
C. Contreras\inst{3,4}
\and
C. Gall\inst{7}
\and
J. Hjorth\inst{7}
\and
C. Ashall \inst{8}
\and
C. R. Burns\inst{9}
\and 
L. Busta\inst{4}
\and
A. Campillay\inst{4}
\and
S. Castell\'{o}n\inst{4}
\and
C. Corco\inst{4}
\and
S. Davis\inst{8}
\and
L. Galbany\inst{10}
\and
C.  Gonz\'{a}lez\inst{4}
\and
S. Holmbo\inst{3}
\and
E. Y. Hsiao\inst{8}
\and
J. R. Maund\inst{11}
\and 
M. M. Phillips\inst{4}
}

\institute{National Astronomical Observatory of Japan, National Institutes of Natural Sciences, 2-21-1 Osawa, Mitaka, Tokyo 181-8588, Japan \\
\email{takashi.moriya@nao.ac.jp}
 \and
School of Physics and Astronomy, Faculty of Science, Monash University, Clayton, Victoria 3800, Australia
\and
Department of Physics and Astronomy, Aarhus University, Ny Munkegade 120, DK-8000 Aarhus C, Denmark
\and
Las Campanas Observatory, Carnegie Observatories, Casilla 601, La Serena, Chile
\and
The George P. and Cynthia Woods Mitchell Institute for Fundamental Physics and Astronomy, Texas A\&M University, College Station, TX 877843, USA
\and
Department of Physics and Astronomy, Texas A\&M University, College Station, TX 77843, USA
\and
DARK, Niels Bohr Institute, University of Copenhagen, Lyngbyvej 2, 2100, Copenhagen \O, Denmark
\and
Department of Physics, Florida State University, Tallahassee, FL 32306, USA
\and
Observatories of the Carnegie Institution for Science, 813 Santa Barbara St, Pasadena, CA, 91101, USA
\and 
Departamento de F\'isica Te\'orica y del Cosmos, Universidad de Granada, E-18071 Granada, Spain
\and
Department of Physics and Astronomy, University of Sheffield, Hicks Building, Hounsfield Road, Sheffield, S3 7RH, UK
}

\date{Received April 08, 2020; accepted July 06, 2020}

 
  \abstract{
  We present optical and near-infrared photometry and spectroscopy of the Type~IIn supernova, (SN) 2014ab, obtained by the Carnegie Supernova Project II (CSP-II) and initiated immediately after its optical discovery. We also present mid-infrared photometry obtained by the \textit{Wide-field Infrared Survey Explorer (WISE)} satellite extending from 56~days prior to the optical discovery to over 1600 days.  The light curve of SN~2014ab evolves slowly, while the spectra exhibit strong emission features produced from the interaction between rapidly expanding ejecta and dense circumstellar matter. The light curve and spectral properties are very similar to those of SN~2010jl.
  The estimated mass-loss rate of the progenitor of SN 2014ab is
  of the order of 0.1~\Msunpyr\
  under the assumption of spherically symmetric circumstellar matter and steady mass loss.
   Although the mid-infrared luminosity increases due to emission from dust, which is characterized by a blackbody temperature close to the dust evaporation temperature ($\sim 2000~\mathrm{K}$), there were no clear signatures of in situ dust formation observed within the cold dense shell located behind the forward shock in SN~2014ab in the early phases. Mid-infrared emission of SN~2014ab may originate from pre-existing dust located within dense circumstellar matter that is heated by the SN shock or shock-driven radiation. Finally, for the benefit of the community, we also  present five near-infrared spectra  of SN~2010jl obtained between 450 to 1300 days post-discovery in the appendix.}

\authorrunning{Moriya, Stritzinger, Taddia, et al.}
\titlerunning{CSP-II observations of the Type~IIn SN~2014ab}

\maketitle

\section{Introduction}
A Type~IIn supernova (SN) exhibits conspicuous, narrow Balmer lines in emission  \citep{schlegel1990iin} formed as a result of the interaction of rapidly expanding SN ejecta with circumstellar matter (CSM, e.g., \citealt{chugai1994iin}). 
 Progenitors of SNe~IIn experience significant mass loss during their pre-SN evolution, leading to a local environment that contains dense CSM \citep[e.g.,][]{ofek2013iinprec,ofek2014iinpre,fraser2013sn2011ht}.
The direct detection of SNe~IIn  progenitors in  pre-explosion imaging \citep{gal-yam2009sn05glprog,prieto2008sn08s} and environmental studies around SNe~IIn 
\citep{anderson2012env,habergham2014intenv,taddia2015iinmetal,galbany2018pisco} 
reveals that a range of masses, including both relatively low-mass progenitors such as red supergiants and high-mass progenitors such as luminous blue variables, serve as the progenitors of  SNe~IIn.

SNe~IIn exhibit a heterogeneous light-curve (LC) evolution \citep[e.g.,][]{nyholm2019ptfiin,taddia2013iin,taddia2015iinmetal,stritzinger2012iin,kiewe2012iin}.
Some SNe~IIn have long-lasting emission and their declines are well-fit by a single power-law  (cf. SN~1988Z; \citealt{stathakis1991sn1988z,turatto1993sn1988z}).  Others  exhibit  rapid, exponentially declining LCs (cf. SN~1998S; \citealt{fassia2000sn1998Sphoto}).  Some SN~IIn have slowly evolving, Gaussian-shaped LC peaks   (cf. SN~2006gy; \citealt{smith2007sn2006gy}), while other SN~IIn exhibit LCs with a  plateau that is akin to  SNe~IIP (cf. SN~1994W; \citealt{sollerman1998sn1994w}). Several SNe~IIn exhibit multiple luminosity peaks \citep{stritzinger2012iin,nyholm2017bump} and some even have documented rise times in excess of   100~days \citep{miller2010sn2008iy,moriya2019hsc16aayt}. 
This LC diversity largely originates from   differences in the properties of the progenitors and CSM  \citep[][]{moriya2013iin,moriya2013sn06gy,chatzopoulos2013chi2,ofek2014ApJ...788..154Ointpeak,dessart2016iin,tsuna2019,takei2019}, compounded by   viewing angle effects  \citep[e.g.,][]{suzuki2019}.

Another important characteristic of SNe~IIn is that they often exhibit a number of observational properties linked to the presence of dust. The dust in SNe~IIn is revealed by the infrared (IR) luminosity evolution \citep[e.g.,][]{fox2011iin,fox2013iin,szali2019spitzer}, as well as by the evolution of prominent emission features \citep[e.g.,][]{pozzo2004sn1998s,taddia2020}. The dust in SNe~IIn has been well-traced in SN~2010jl. SN~2010jl is a slowly-declining 1988Z-like object that exhibits a power-law luminosity decline at early times \citep{stoll2011sn2010jl,zhang2012sn2010jl,maeda2013sn2010jl,fransson2014sn2010jl,ofek2014sn2010jl,jencson2016sn10jl}.
Although SN~2010jl was observed intensively over a range of wavelengths \citep{chandra2012sn2010jl,chandra2015sn2010jl,williams2015sn2010jl,bevan2020}, the origin of the dust traced by its IR observations is still a matter of debate. The dust might have  formed within a cool dense shell located within the wake of the supernova shock wave which is positioned at the interface between the rapidly expanding supernova ejecta and the dense CSM \citep[e.g.,][]{smith2011sn10jl,gall2014sn2010jl,maeda2013sn2010jl}. Alternatively,   the IR emission could be attributed to  pre-existing dust located around the progenitor \citep{andrews2011sn2010jl,fransson2014sn2010jl,saragni2018sn2010jl,chugai2018sn2010jl}. The possibility that most IR emission observed in SNe~IIn originates from pre-existing dust has also been proposed \citep[e.g.,][]{fox2011iin}. To clarify the origin of dust in SNe~IIn, more detailed observations of SNe~IIn covering a wide wavelength range are required.

In this paper, we report on the observations of the long-lasting (1988Z-like) Type~IIn SN~2014ab, initiated by the Carnegie Supernova Project-II (CSP-II; \citealt{2019PASP..131a4001P,2019PASP..131a4002H}). Although this object  spectroscopically resembles the well-observed SN~2010jl, it does exhibit different LC evolution and dust signatures. 
SN~2014ab offers\ a new and important clue on the nature of slowly-declining SNe~IIn, including details about the dust in their nearby environment. 

\begin{figure}
\centering
\includegraphics[width=1\linewidth]{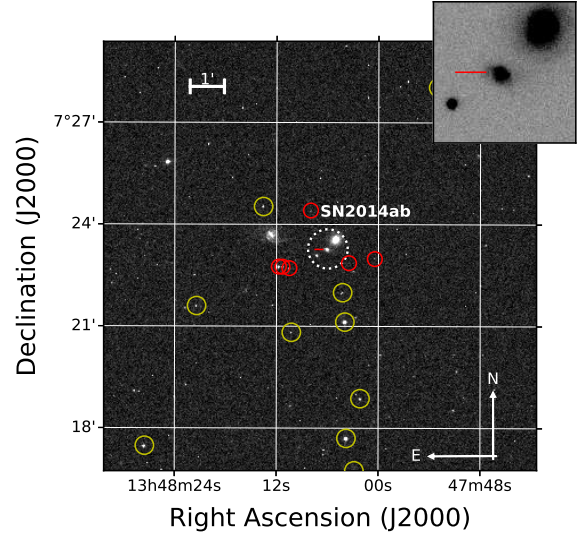} 
\caption{Finding chart  constructed using a single $r$-band image obtained on 03 March 2014 with the Swope 1.0-m telescope. The position of  SN~2014ab is indicated with a dotted circle and the approximate area contained within the circle is shown in the upper right inset
Optical and NIR local sequence stars are indicated with yellow and red circles, respectively.}
\label{fig:FC}%
\end{figure} 

\input{tables/specjournal.tex}

\section{SN 2014ab: discovery, distance, and reddening}
\label{sec:host-galaxy} 
The Catalina Real-time Transient Survey (CRTS) reported the discovery of a transient with an apparent magnitude of 16.4 in MCG +01-35-037 \citep{2014CBET.3826....1H}. 
First designated PSN~J13480599+0723164 when it was posted to  the web page of the  Central Bureau for Astronomical Telegrams, the transient was identified by CRTS in an image  taken by the  Catalina Sky Survey (CSS)  on March 9.43 UT (i.e., JD-2456725.93). With coordinates
R.A. (J2000.0) $=$ 13h48m05s.99 and Decl. (J2000.0) $= +07^\circ$23$'$16$\farcs$40, the transient (aka SNhunt237) is located  1\farcs5 north of the host galaxy (see Fig.~\ref{fig:FC}). 

Within 24 hours  of the discovery (Mar. 10.25 UT), the CSP-II obtained a near-IR (NIR) spectrum with the Folded-port Infrared Echellette (FIRE; \citealt{simcoe2013FIRE}) spectrograph attached to the  Magellan Baade 6.5-m Telescope located at Las Campanas Observatory (LCO). 
This spectrum revealed an SN~IIn and upon the reporting  of its spectral classification, the transient obtained the International Astronomical Union (IAU) designation of SN~2014ab.
The CSP-II classification was promptly confirmed by PESSTO (Public ESO Spectroscopic Survey for Transient Objects; \citealt{smartt2015pessto}) based on a  visual-wavelength spectrum obtained with the New Technology Telescope   (+ EFOSC; \citealt{buzzoni1984efosc}), located at the ESO La Silla Paranal Observatory \citep{2014ATel.5968....1F}.

According to the NASA Extracgalactic Database (NED), the host-galaxy MCG +01-35-037 has a redshift $z=0.02352$. 
Turning to the medium-dispersion spectra of SN~2014ab, obtained with the ESO La Silla Paranal Observatory's Very Large Telescope (VLT) equipped with the X-shooter (see below),
we obtain 
an averaged redshift of $z=0.02262\pm 0.00001$ from the narrow H$\alpha$ emission component in the three spectra.
Adopting $H_0 = 73.2\pm2.3$ km~s$^{-1}$~Mpc$^{-1}$ \citep{2018ApJ...869...56B}, $\Omega_M = 0.27$, and $\Omega_{\Lambda} = 0.73$,  this corresponds to
a luminosity distance of $94.3\pm6.6$ Mpc, or a distance modulus $\mu = 34.87\pm0.15$ mag.

According to \citet{2011ApJ...737..103S} and reported in NED\footnote{\url{https://ned.ipac.caltech.edu/}},  the Milky Way visual-band extinction in the direction of SN~2014ab is  $A^{MW}_V = 0.083$~mag. This value is inferred from infrared-based dust maps and assumes a \citet{1999PASP..111...63F} reddening law with $R_V = 3.1$.

An inspection of the visual-wavelength spectra of SN~2014ab presented below reveals no \ion{Na}{i}~D features at the redshift of the host galaxy and there is no evidence of the 5780~\AA\ diffuse interstellar band. We assume, therefore, in the following that SN~2014ab suffers from minimal to no host-galaxy reddening.

\section{Observations}

\subsection{\textit{WISE} mid-infrared photometry}

We include ten epochs of two-channel mid-IR (MIR) imaging of SN~2014ab  serendipitously  obtained by the \textit{Wide-field Infrared Survey Explorer} (\textit{WISE}) satellite \citep{wright2010wise,mainzer2014neowise}, 
and  downloaded from  NASA/IPAC Infrared Science Archive\footnote{\url{https://irsa.ipac.caltech.edu/}}. 
Photometry retrieved from the archive includes  \textit{W1}-band measurements at $3.4~\mu\mathrm{m}$ and \textit{W2}-band measurements at $4.6~\mu\mathrm{m}$. 
This photometry was previously presented by \citet{jiang2019wise}. The first detection by \textit{WISE} is at JD 2456671.59, which is 56~days before the optical discovery at JD 2456725.93 in the observer frame. 
The \textit{WISE} photometry is listed in Table~\ref{wisephoto}.

\begin{figure}
\centering
\includegraphics[width=0.95\columnwidth]{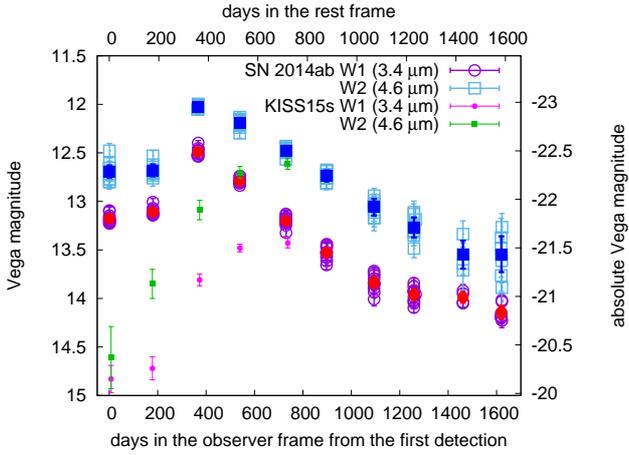} 
\caption{Mid-infrared light curves of SN~2014ab obtained with the \textit{WISE} satellite. The open symbols show the raw data and the filled symbols show co-added magnitudes of the nearby epochs. We also present the photometry of KISS15s, which is another SN~IIn with serendipitous \textit{WISE} observations \citep{kokubo2019kiss15s}. The KISS15s photometry is plotted with the right and top axes. The time origin in KISS15s is its optical discovery.
}
\label{fig:wise}%
\end{figure} 

\subsection{Carnegie Supernova Project-II observations of SN~2014ab}

The CSP-II obtained broad-band optical and NIR imaging with facilities located at LCO. In particular, twenty-two epochs of optical ($ugriBV$) imaging was taken with the Henrietta Swope 1.0-m telescope equipped with a direct optical camera containing an e2V CCD, while five epochs of NIR images were taken with the Ir\'en\'ee du Pont 2.5-m telescope equipped with RetroCam \citep{hamuy2006}. 
The $ugriBV$-band photometry  extends, relative to the first \textit{WISE} detection, from  $+$56~d to $+$485~d, with a significant gap between $+$140~d to $+$360~d when SN~2014ab was located behind the Sun. The NIR photometry is more limited in coverage, ranging  between  $+56$~d to $+121$~d. All dates presented here are in the observer frame.

The CSP-II data were reduced to photometry following standard prescriptions that are well documented in \citet[][and references therein]{2017AJ....154..211K}. Prior to computing the photometry of the SN from each science image, deep optical and NIR templates of the host were obtained with the du Pont telescope equipped with a direct CCD camera  and with the  Magellan Baade telescope equipped with the NIR imager FourStar \citep{persson2013fourstar}. 

PSF photometry of SN~2014ab was computed from the template-subtracted science images and its instrumental magnitudes were calibrated  relative to a local sequence of stars located in the field of the host.  
The optical local sequence consists of twenty-three stars and the NIR local sequence consists of six stars. 
The optical local sequence was calibrated relative to standard star fields observed  over a minimum of three photometric nights, while the NIR sequence were calibrated relative to \citet{1998AJ....116.2475P} ($JH$ band) and \citet{2017AJ....154..211K} ($Y$ band) standard star fields observed over four and three photometric nights, respectively.   Optical and NIR photometry of the local sequence stars in the  "standard" photometric systems are listed in Table~\ref{opticallocseq} 
and Table~\ref{nirlocseq}, respectively. Finally, the optical and NIR photometry of SN~2014ab in the natural system are listed in Table~\ref{photometry}.

The CSP-II obtained  a single visual-wavelength spectrum of SN~2014ab with the du Pont telescope equipped with Wide Field Re-imaging CCD Camera (WFCCD) and four epochs of NIR spectra with the Magellan Baade telescope equipped with the FIRE  spectrograph. In addition to these data, we obtained optical/NIR spectra from the VLT  (+ X-shooter, \citealt{vernett2011xshooter}; Program ID: 092.D-0645, PI Gall), along with a single NTT (+ EFOSC) spectrum reported by PESSTO.
A journal of the spectroscopic observations is provided in Table~\ref{specjor}.
The X-shooter data were obtained in nodding (ABBA) mode and underwent basic reduction steps in "stare" mode for the UVB and VIS arms and "nodding" mode for the NIR arm using the EsoReflex workflow \citep{2013A&A...559A..96F}. The UVB and VIS single-stare frames were combined using custom python scripts\footnote{\url{https://github.com/jselsing/xsh-postproc}} (\citealt{2019A&A...623A..92S}) and optically extracted, slit loss corrected, and corrected for heliocentric velocity using a custom IDL programs. The telluric correction was performed by using \texttt{Molecfit} \citep{2015A&A...576A..77S,2015A&A...576A..78K}.
The other spectroscopic data  were reduced following standard procedures as previously described  by \citet{hamuy2006} (optical) for  and \citet{2019PASP..131a4002H} (NIR). 
Finally, the  flux scale of each of the one-dimensional, visual-wavelength spectra were adjusted such that their synthetic colors  match the more accurate broad-band colors computed from  photometry \citep[see, e.g.,][]{2014PASP..126..324B}.

In addition to the spectra of SN~2014ab, we also present CSP-II spectra of SN~2010jl in Appendix~\ref{sec:10jl}.
As  demonstrated below, SN~2014ab shares some common features with SN~2010jl.  These spectra were reduced following the same manner as applied to the data of SN~2014ab.

\section{Results}

\subsection{Photometry and $B-V$ color-curve evolution}

Plotted in Fig.~\ref{fig:wise} are the two-channel MIR LCs computed from  \textit{WISE} observations  covering  over 1600 days of evolution. The object reached a maximum between 200 and 400 days post-detection, followed by a slow decline ($\sim 0.0015~\mathrm{mag~day^{-1}}$) over the next 1200 days. The object appears to have reached a \textit{W1}-band peak brightness of $\leq -22.5$ mag and the \textit{W2}-band reached a peak  of $\leq -23$ mag.  

\begin{figure}
\centering
\includegraphics[width=0.95\columnwidth]{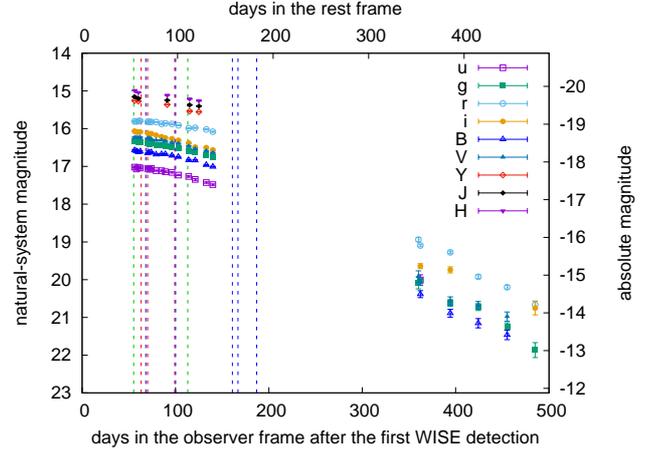} 
\caption{Multi-color, optical ($ugriBV$) and NIR ($YJH$)  light curves of SN~2014ab obtained by the CSP-II. The data have been corrected for Galactic extinction. Vertical lines indicate epochs that optical (green), NIR (red), or optical+NIR (blue) spectroscopic observations were obtained.
}
\label{fig:lightcurves}%
\end{figure}

The  CSP-II optical and NIR LCs of SN~2014ab  are  plotted in Fig.~\ref{fig:lightcurves}. 
The optical photometry extends from $+$56~d to $+$485~d relative to the first \textit{WISE} detection in the observer frame, while the NIR photometry extends between +56~d to +121~d.
The  photometry shown in the figure has been corrected for Galactic reddening, and placed on the absolute magnitude scaling adopting the distance computed in Section~\ref{sec:host-galaxy}. The optical LCs at the time of discovery range from $-17.5$ mag to $-19$ mag, while the NIR absolute magnitudes reach around $-20$ mag.
 In the initial phases of our follow-up, the optical/NIR LCs slowly evolve  until $+$123~d when follow-up ceased for $\sim$ 200 days. For example, the $g$-band LC declines 0.36 mag per 100 days. Follow-up then continued at around $+$350~d and extended until around $+$480~d. During this time the $g$-band LC decline rate was 1.4 mag per 100 days.

\begin{figure}
\centering
\includegraphics[width=0.95\columnwidth]{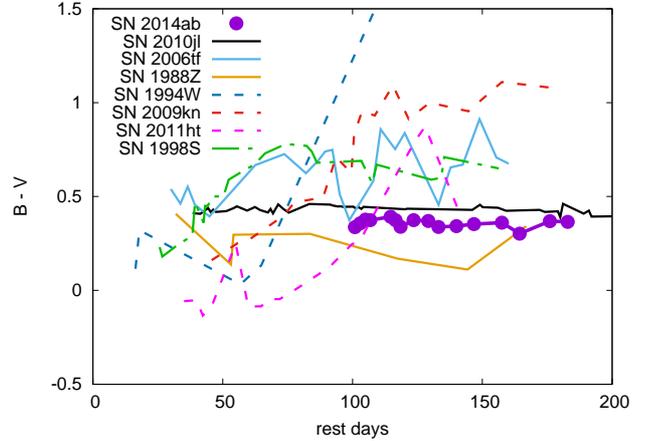}
\caption{Observed $B-V$ color curve of SN~2014ab compared with those of three different SNe~IIn subtypes. The reddening is corrected based on the values reported in the literature.
The explosion date, i.e., day zero, for SN~2014ab is arbitrarily set to be 100~days before the optical discovery, or 46~days before the \textit{WISE} first detection in the rest frame. 
The comparison sample including objects of the slowly-declining 1988Z-like subtype (solid lines) including SN~2010jl \citep{fransson2014sn2010jl}, SN~2006tf \citep{smith2008sn2006tf}, and SN~1988Z \citep{turatto1993sn1988z}, as well as the 1994W-like plateau subtype (dashed lines) including SN~1994W \citep{sollerman1998sn1994w}, SN~2009kn \citep{kankare2012sn2009kn}, and SN~2011ht \citep{mauerhan2013sn2011ht}, and finally, the prototypical rapidly-declining SN~1998S (dot-dashed line, \citealt{fassia2000sn1998Sphoto}).
}
\label{fig:BmVcolor}%
\end{figure}

Figure~\ref{fig:BmVcolor} presents the $B-V$ color evolution of SN~2014ab and other representative SNe~IIn. SN~2014ab has little color evolution during our observations lasting for about 100~days. A lack of  significant color evolution appears to be a hallmark of other slowly declining SNe~IIn such as SN~1988Z and SN~2010jl. In particular, the color-curve evolution of  SN~2010jl is very much like that of SN~2014ab. As discussed later, the spectra of SN~2014ab  are also very similar to those of SN~2010jl.

\begin{figure*}
\centering
\includegraphics[width=1.8\columnwidth]{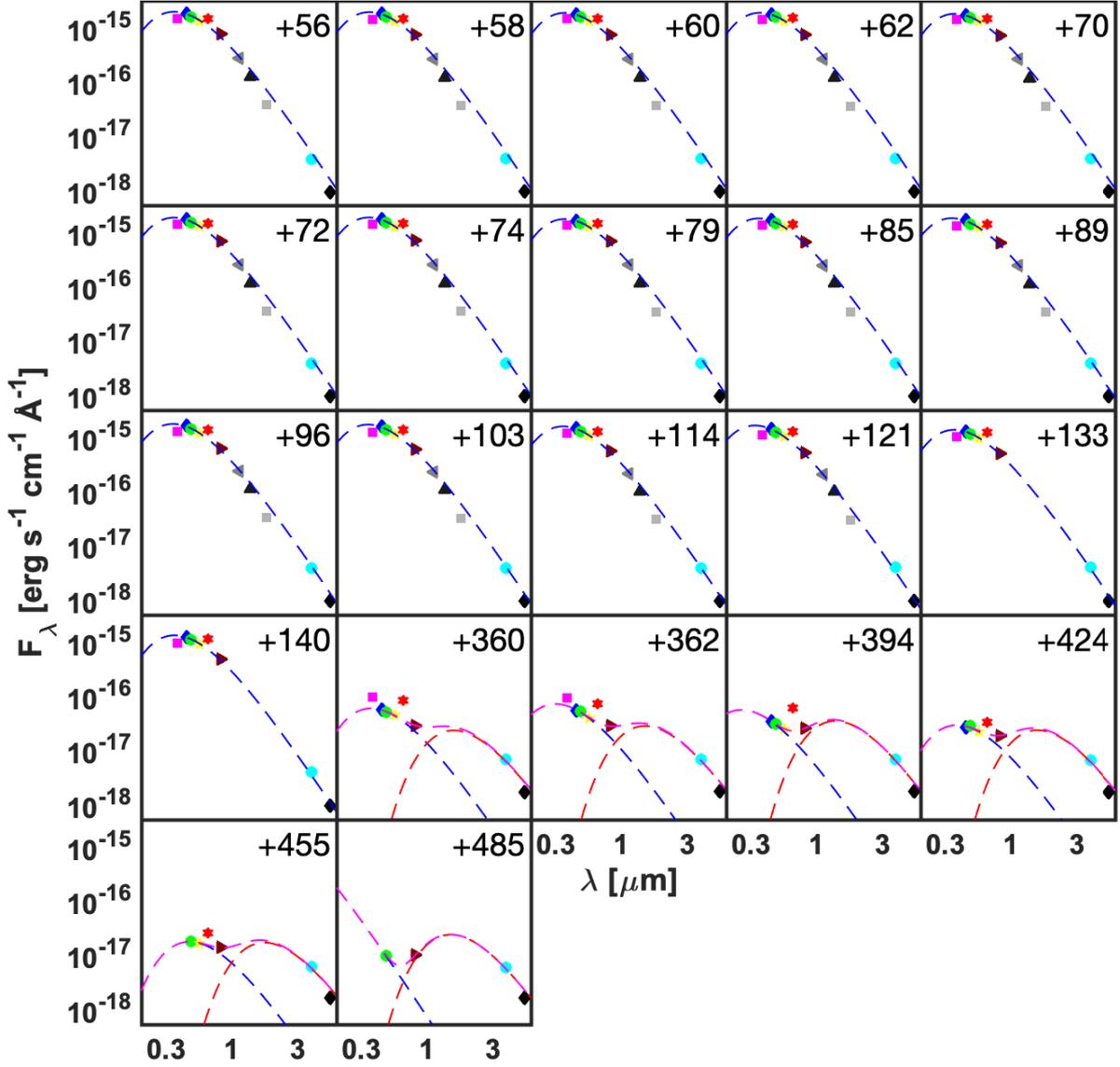} 
\caption{Spectral energy distributions (SEDs) of SN~2014ab.  The first 16 quadrants from the top contain SEDs constructed from flux points covering optical, NIR, and MIR wavelengths and extend from +56~d to +140~d relative to the time of first \textit{WISE} detection. Each of these  SEDs are fit with a single black body (BB) function, where the optical flux is dominating. We excluded $u$ and $r$ band from the fit due to line blanketing and the strong H$\alpha$ emission line, respectively. The SEDs plotted in the other quadrants are constructed from optical and MIR flux points and extend from $+$360~d to $+$485~d relative to first \textit{WISE} detection. The late-phase SEDs clearly exhibit an excess of flux relative to a single BB fit to the optical flux points. We therefore fit the late SEDs with a two-component BB function.   In the last quadrant the optical BB fit is not reliable as only $g$ and $i$ are available.
}
\label{fig:SEDs}%
\end{figure*} 

\begin{figure}
\centering
\includegraphics[width=0.95\columnwidth]{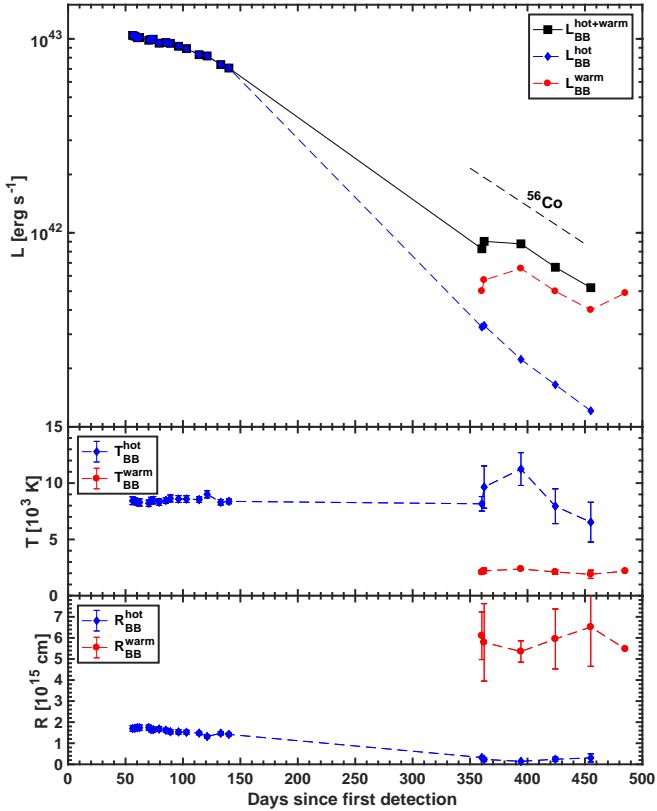} 
\caption{Bolometric LC (top) and the BB-temperature (middle) and BB-radius (bottom) profiles of SN~2014ab, computed from  fitting single and double BB functions to the SEDs of SN~2014ab. 
}
\label{fig:UVOIR}%
\end{figure} 

\begin{figure*}
\centering
\includegraphics[width=1.8\columnwidth]{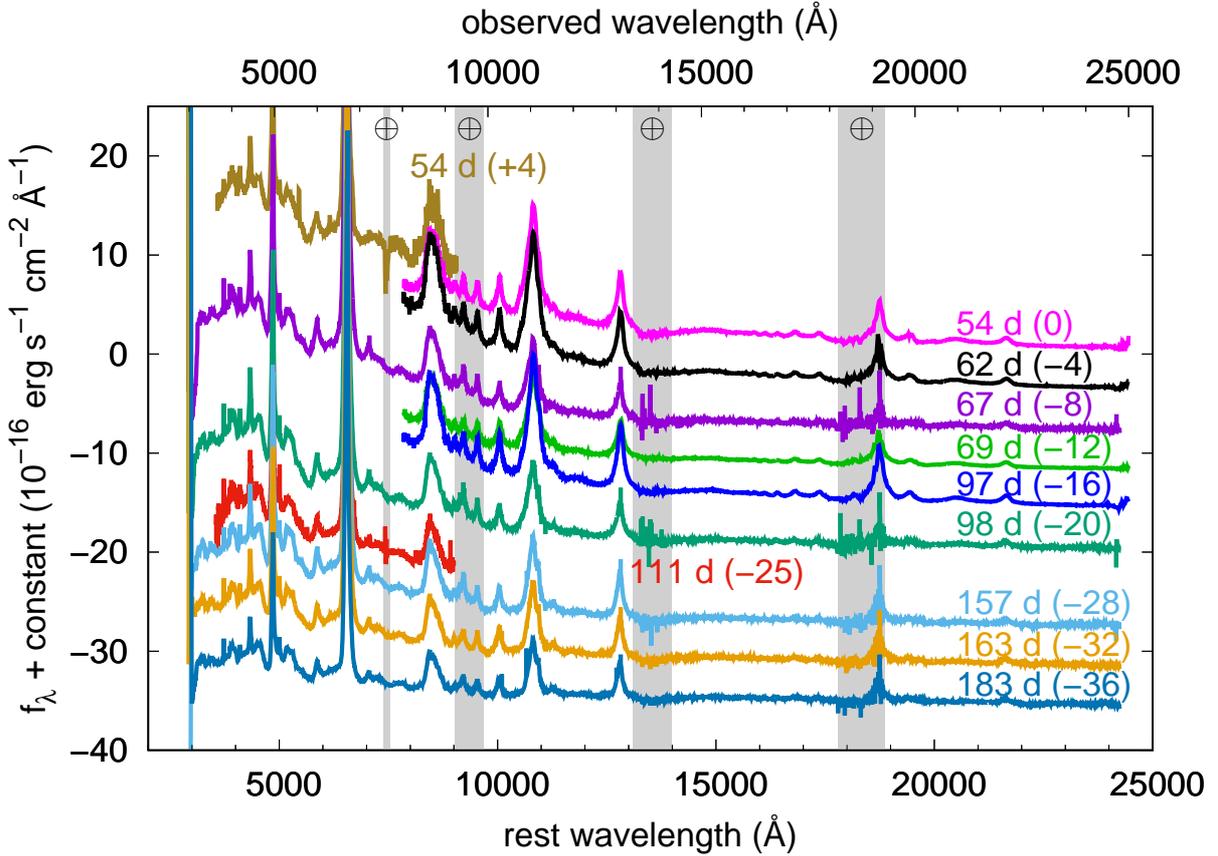} 
\caption{Optical through NIR wavelength spectroscopy of SN~2014ab, extending from +54~d to 183~d (restframe days) past the first \textit{WISE} detection. The numbers in the parentheses indicate the flux offset adopted for the spectra in the unit of $10^{-16}$~erg~s$^{-1}$~cm$^{-2}$~\AA$^{-1}$. The shaded wavelength regions indicate areas of strong telluric absorption. The last three spectra are not calibrated by photometry because of the lack in the photometric information.}
\label{fig:allspectra}%
\end{figure*}

\begin{figure}
\centering
\includegraphics[width=0.95\columnwidth]{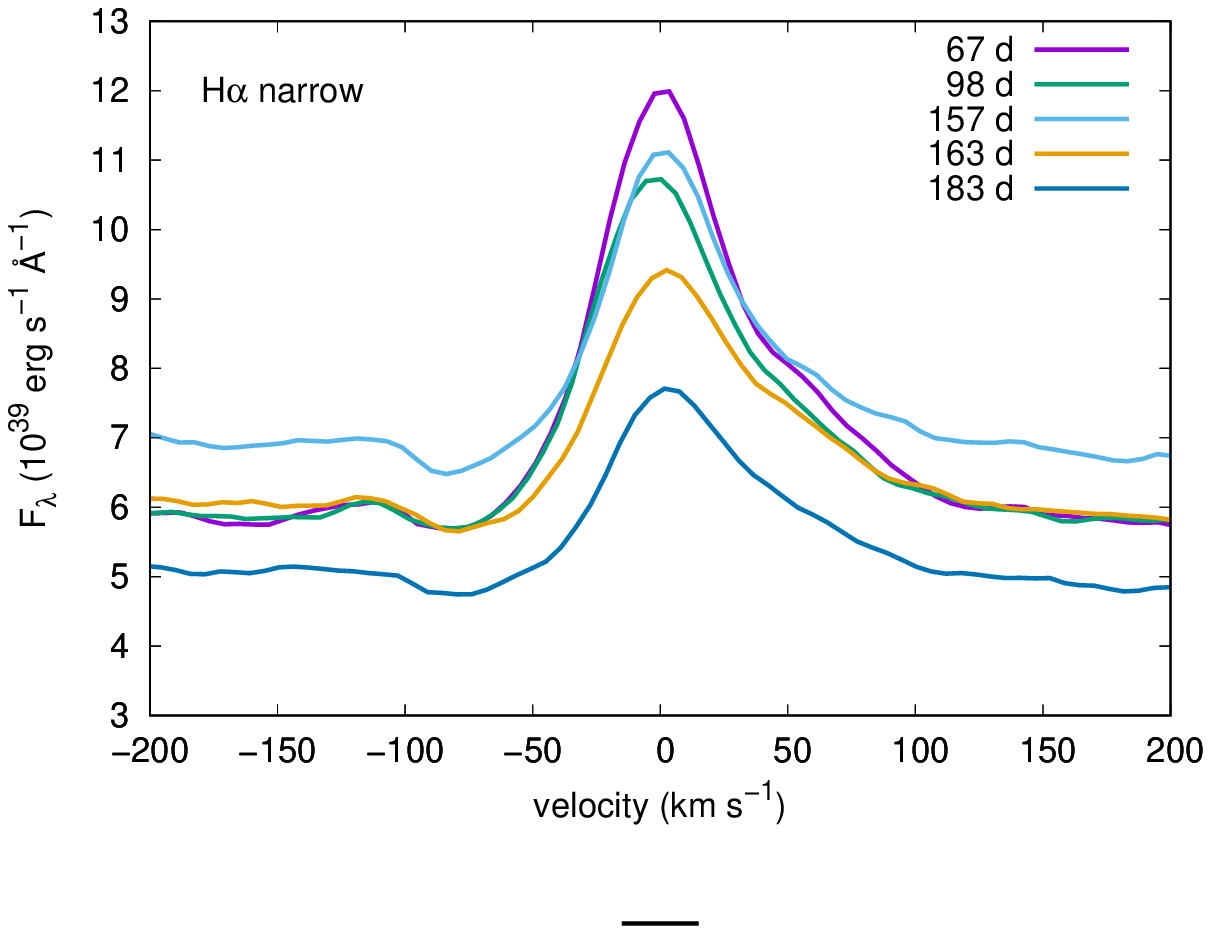} 
\includegraphics[width=0.95\columnwidth]{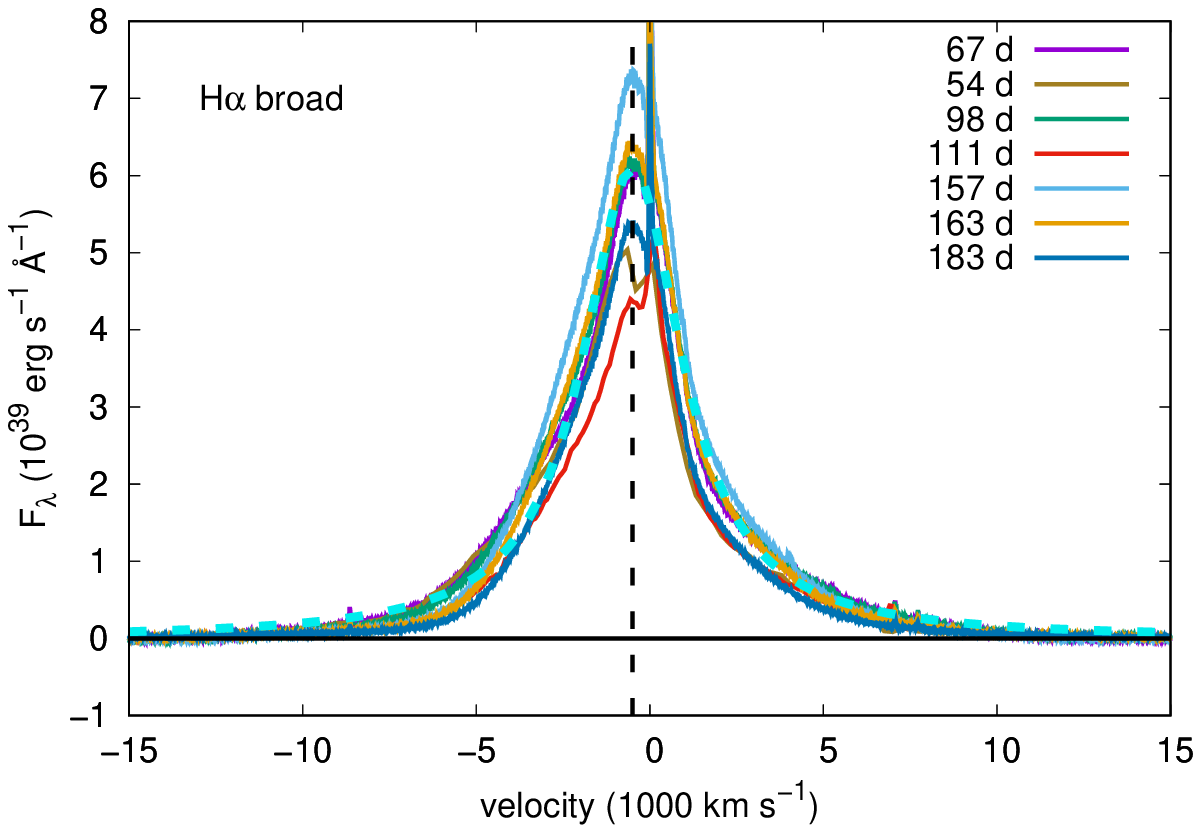} 
\includegraphics[width=0.95\columnwidth]{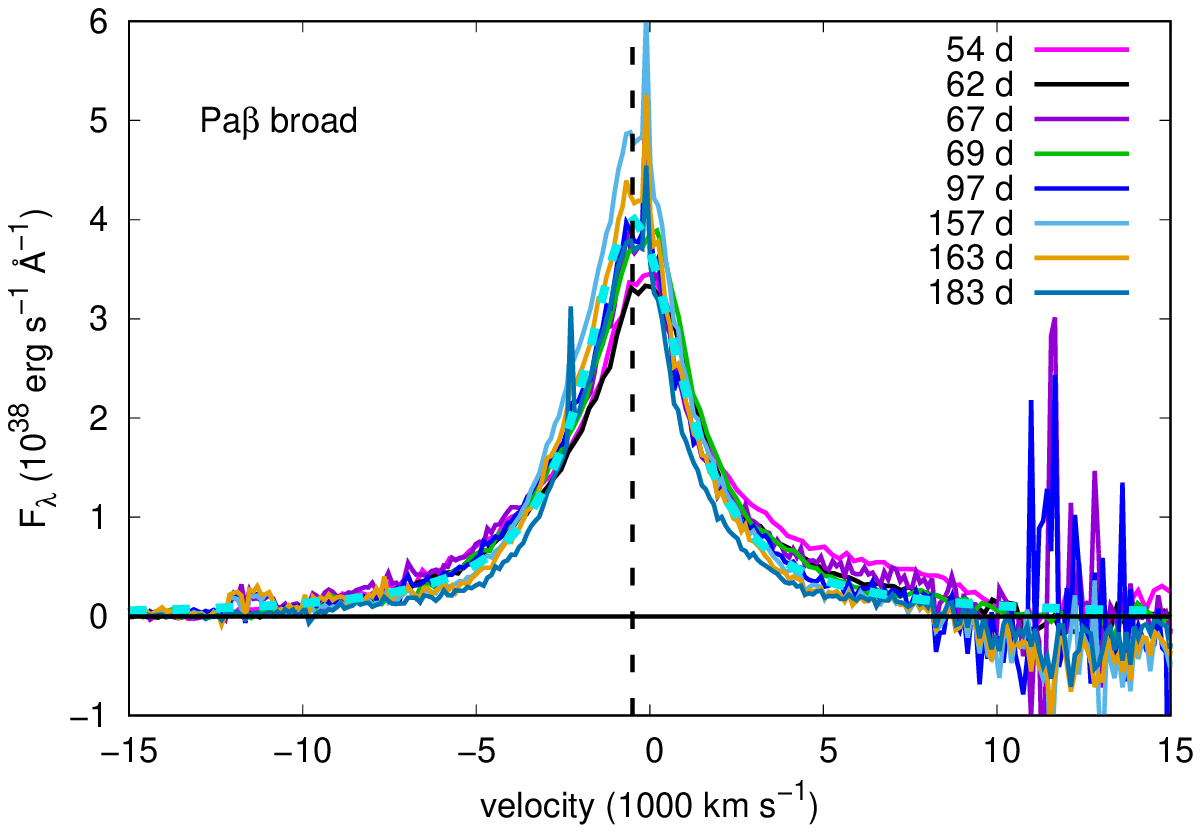} 
\caption{Time-evolution of the narrow- (top panel) and broad-component  (middle panel) H$\alpha$ profiles, and the broad-component (bottom panel) of the Pa-$\beta$ profile, plotted in velocity space. Over-plotted as dashed cyan lines are Lorentzian functions with the FWHM velocity of 3500~\kmps\ and the peak shifted by $-500~\kmps$.
The redshift is determined by the average of the narrow emission peak.
Both broad components are overall symmetric and   do not exhibit significant evolution over the duration of our followup observations.
}
\label{fig:hprofiles}%
\end{figure} 

\begin{figure}
\centering
\includegraphics[width=0.95\columnwidth]{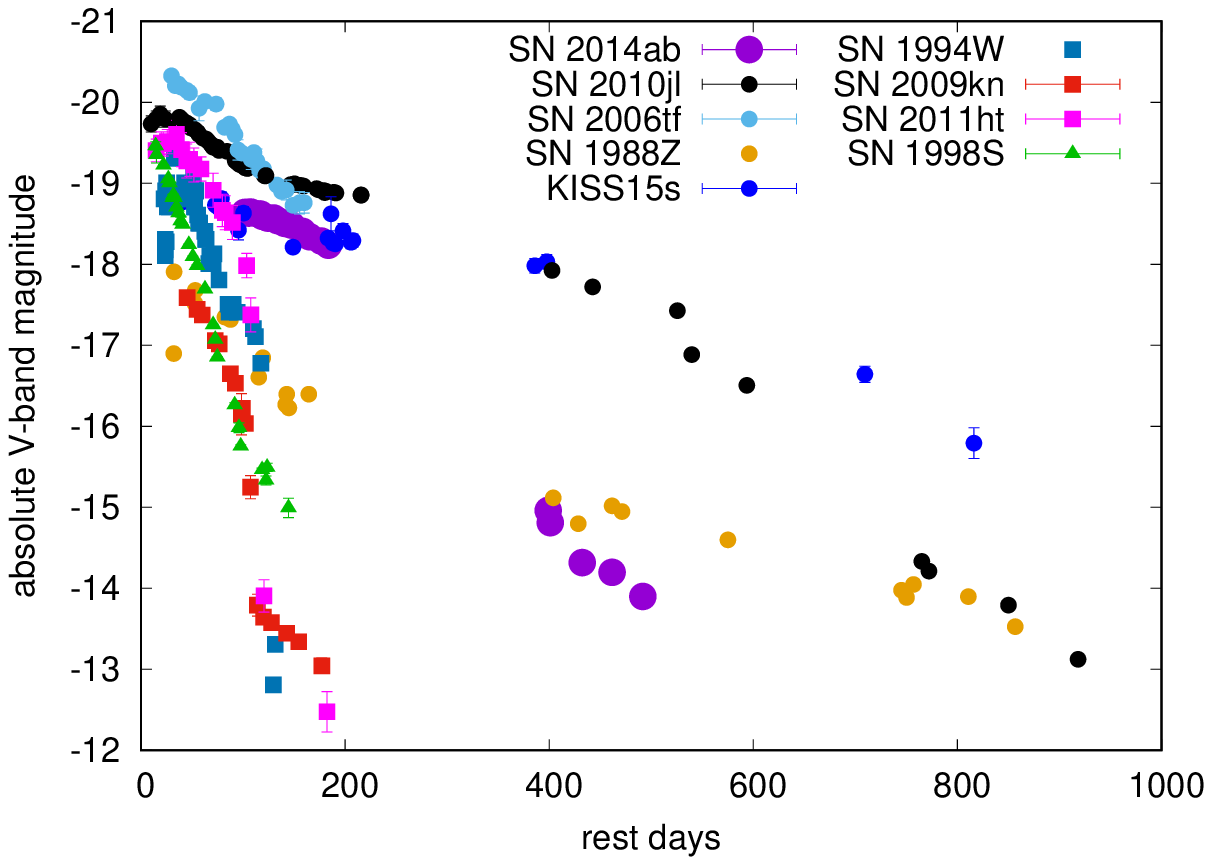} 
\caption{Absolute $V$-band LC of SN~2014ab compared with well-observed examples of the three subtypes of SN~IIn discussed by \citet{taddia2013iin}.
This includes the long-lasting 1988Z-like objects (circles) SN~1988Z \citep{turatto1993sn1988z}, SN~2006tf \citep{smith2008sn2006tf}, SN~2010jl \citep{stoll2011sn2010jl,zhang2012sn2010jl,maeda2013sn2010jl,fransson2014sn2010jl}, and KISS15s \citep{kokubo2019kiss15s},   the 1994W-like objects (squares) including SN~1994W \citep{chugai2004sn94w}, SN~2009kn \citep{kankare2012sn2009kn}, and SN~2011ht \citep{mauerhan2013sn2011ht}, which show a LC pleateau, and finally, the prototypical, rapidly-declining SN~1998S (triangles, \citealt{fassia2000sn1998Sphoto}).
}
\label{fig:lightcurvecomparison}%
\end{figure} 

\begin{figure}
\centering
\includegraphics[width=0.95\columnwidth]{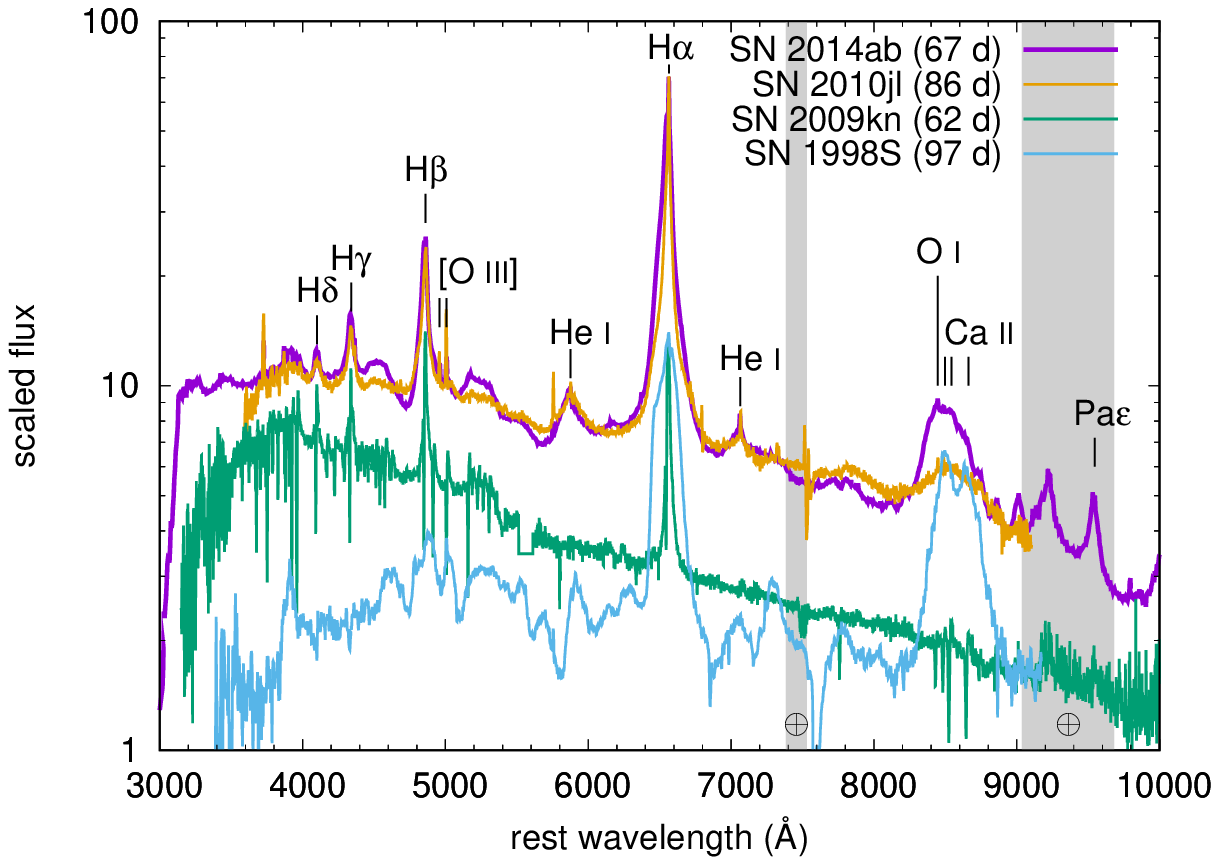} 
\includegraphics[width=0.95\columnwidth]{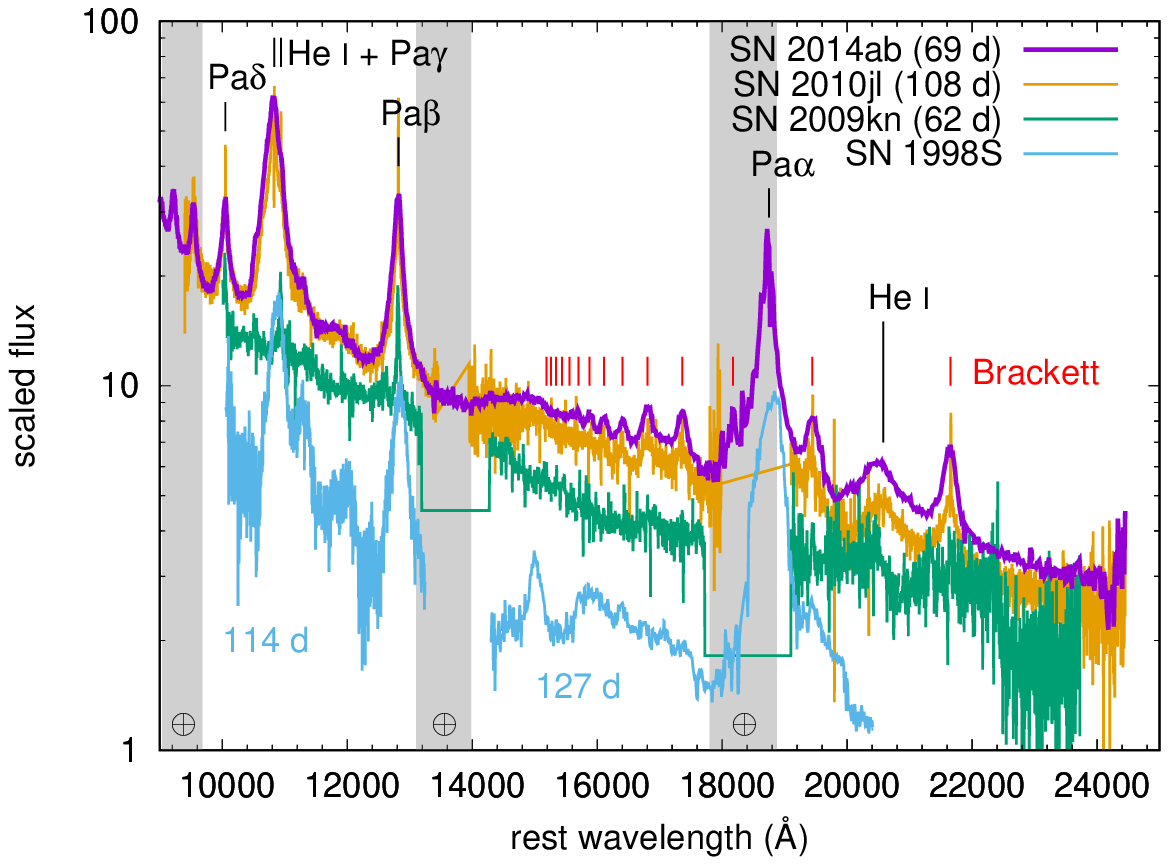} 
\caption{Comparison of our +67~d visual-wavelength (top) and NIR-wavelength (bottom) spectrum of SN~2014ab  with other  SN~IIn subtypes. Comparison objects are the slowly declining SN~2010jl \citep{borish2015sn2010jl}, the 1994W-like SN~2009kn \citep{kankare2012sn2009kn}, and the rapid-declining SN~1998S  \citep{fassia2001sn1998sspec}. Prominent ions are indicated and labeled. Regions  suffering from strong telluric lines  are shaded. Note that the NIR spectrum of SN~1998S is composed of data taken on two different but close in time epochs as indicated. Note that the previously unpublished visual-wavelength spectrum of SN~2010jl was obtained by the CSP-II (see Appendix~\ref{sec:10jl} for details). }
          \label{fig:speccompandid}%
\end{figure}

\begin{figure}
\centering
\includegraphics[width=0.95\columnwidth]{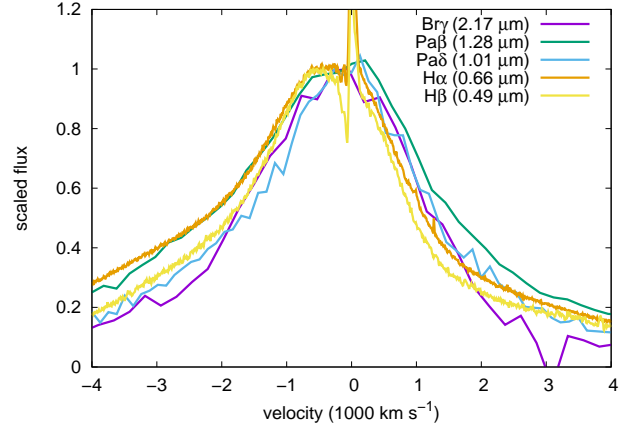} 
\caption{
Peak of the hydrogen emission lines. We use the optical spectrum at $+67$~d and the NIR spectrum at $+69$~d. The peak of the broad component does not show a wavelength-dependent shift expected if dust is formed at the emission line forming region, i.e., the emission lines at the shorter wavelengths to have more blue-shifted peaks. The flux is normalized to match the emission peak.
}
\label{fig:hemissions}%
\end{figure} 

\begin{figure}
\centering
\includegraphics[width=0.95\columnwidth]{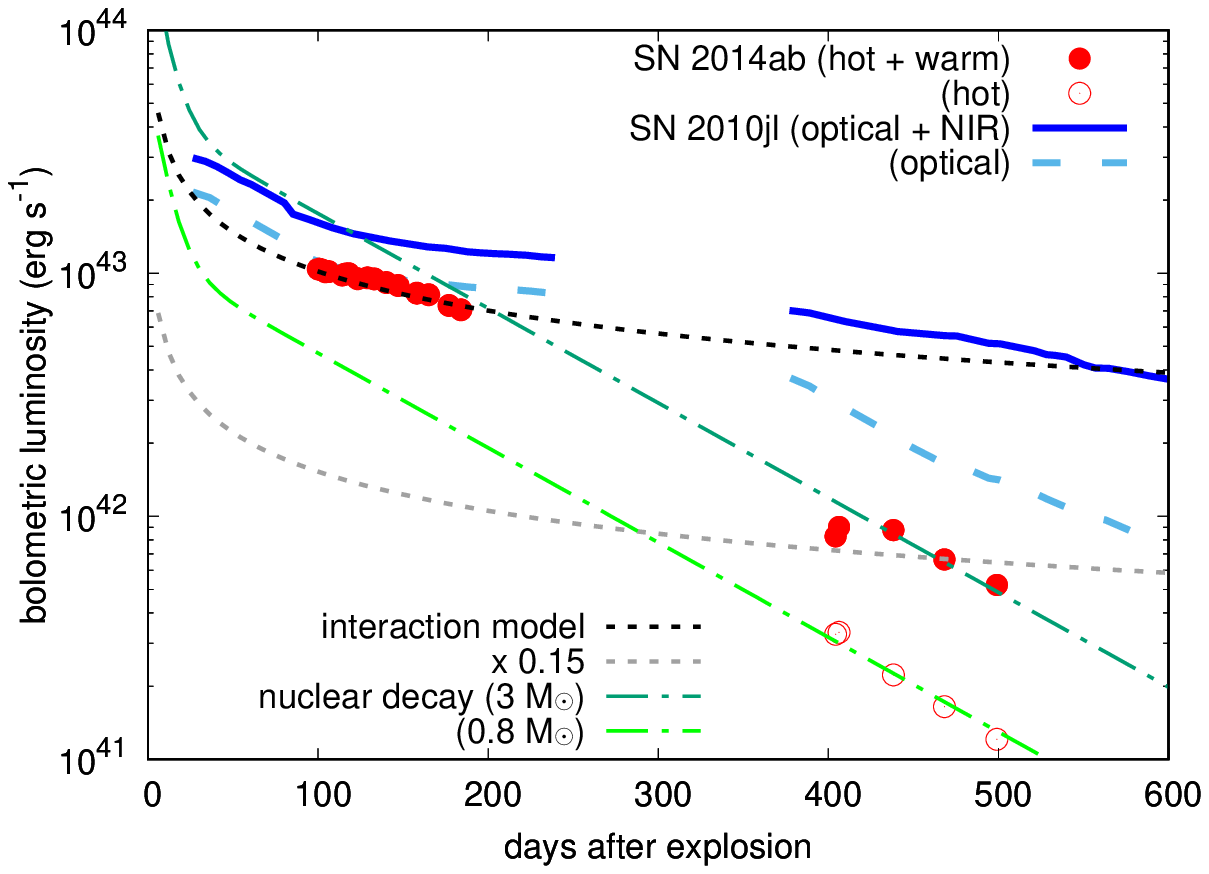} 
\caption{
Bolometric LC of SN~2014ab compared to that of SN~2010jl. Here the explosion epoch of SN~2014ab is  assumed to occur one hundred days prior to the optical discovery. The ``hot + warm" luminosity is obtained by adding the two blackbody components and the ``hot" luminosity is obtained only with the optical blackbody component in SN~2014ab. The LCs of SN~2010jl are from \citet{fransson2014sn2010jl}. The "optical + NIR" luminosity is the luminosity in the wavelength range from 3600~\AA\ to 24000~\AA. The "optical" luminosity wavelength range is from 3600~\AA\ to 9000~\AA. We show the power-law bolometric LC fit that is analytically expected for interacting SNe and the total nuclear deposition energy from the decay of \Ni\ and \Co.
}
\label{fig:lightcurvemodels}%
\end{figure}

 \subsection{Spectral energy distributions and UVOIR light curve}
\label{bolometicLC}
Here, we use the broad-band optical through MIR photometry of SN~2014ab to construct spectral energy distributions (SEDs) for each observed epoch. These are then  fit with black-body (BB) functions in order to estimate the BB temperature ($T_{BB}$), BB radius ($R_{BB}$), and BB luminosity ($L_{BB}$) profiles associated with the underlying emission region(s). 

Plotted in Fig.~\ref{fig:SEDs}, we present the SEDs of SN~2014ab, extending from $+$56~d to $+$485~d relative to the first \textit{WISE} detection. To build these SEDs, first the  photometry  at each epoch  was  corrected for reddening and then  each optical and NIR point was placed on  the AB system using the AB offsets given in \citet{2017AJ....154..211K}; see their Table 16.  Then each photometric point was converted to flux following Eq.~F5  of \citet{2017AJ....154..211K}. If there were gaps in the optical or NIR coverage at any particular epoch, they were estimated from neighboring photometry through the use of linear interpolation. The \textit{WISE} photometry, which is given in the Vega system, was converted to flux using the flux zero points given in \citet{wright2010wise}.

Next, each SED was fit with a BB function. When we compute the best-fit BB functions to the SEDs, the $u$- and $r$-band flux points were excluded because they are affected by significant line blanketing and the strong H$\alpha$ emission line, respectively (Section~\ref{sec:spectroscopy}). We find that the obtained BB fit matches well to the continuum of the spectra when available. The BB temperature at the early epochs is around 8000~K.
An inspection of Fig.~\ref{fig:SEDs} reveals that from the $+$360~d SED and onwards, an excess of MIR flux is present relative to expectations of a single BB function fit (blue dashed line). We therefore fit the SEDs extending from $+360$~d to $+484$~d  with an additional  BB-function component, which is plotted as red dashed line, while the combined BB functions are overplotted in magenta.
We refer to the first hot BB component as $BB_{hot}$ and the second warm BB component as $BB_{warm}$.
$BB_{hot}$ remains around 8000~K, although the uncertainty is large at the late epochs.
We cannot tell if $BB_{warm}$ is affected by the presence of emission lines in the MIR bands because we  lack MIR spectra.
We therefore turn to the MIR spectra of the intermediate-luminosity-red-transient NGC~300-2008-OT, which were obtained around 400~days and 600~days
\citep{ohsawa2010ngc300ot}. This gap transient is thought to have been powered by circumstellar interaction \citep{bond2009ot300,berger2009ot300}. These  spectra  reveal no prevalent emission features. The $BB_{warm}$ component fit to the MIR flux points of SN~2014ab suggests  $T_{BB}^{warm} \sim 2000~K$, which is similar to the value inferred  for SN~2010jl ($1500-2000~\mathrm{K}$, \citealt{fransson2014sn2010jl}). We note that $T_{BB}^{warm}$ is mostly constrained by the two MIR photometric bands. We also tried BB fits by forcing $T_{BB}^{warm}=1000~\mathrm{K}$ and found that the BB fit significantly overestimates the \textit{W2} photometry. Therefore, $T_{BB}^{warm}$ would not be far from 2000~K ($\simeq 1500-2000$~K) and is close to the dust evaporation temperature.

Next, by integrating the BB function fits we obtain the UVOIR LC of  SN~2014ab as plotted in the top panel of Fig.~\ref{fig:UVOIR}. Specifically, we plot $L^{hot}_{BB}$ (blue),  $L^{warm}_{BB}$ (red) and the summation of the two (black).  We find that on $+56$~d, SN~2014ab  reached $L^{tot}_{BB} \approx 1.0\times10^{43}$ erg~s$^{-1}$, and by $+360$~d this dropped by an order of magnitude. Furthermore, we find that at $+$360~d that the $L^{warm}_{BB}$ dominates over the emission of $L^{hot}_{BB}$  by a factor of $\sim3$ and this increases to a factor of $\sim 5$ by $+$485~d. Comparison of the $T^{warm}_{BB}$ and $T^{hot}_{BB}$ values reveals that the former characterized  by $T^{warm}_{BB} \sim 2000$~K  is around a factor of $\sim 4$ lower than the latter over the duration covered by our late phase data. Over the same time,  $R^{warm}_{BB}$ is found to be around a factor of $\sim 6$ higher than $R^{hot}_{BB}$.  These attributes suggest that the emission of the $BB_{warm}$ component  is  related to  dust reprocessing of shorter wavelength photons to longer wavelength photons.  

The radiated energy from +56~d to +140~d in Fig.~\ref{fig:UVOIR} is $6.5\times 10^{49}~\mathrm{erg}$ and that from +360~d to +455~d is $6.2\times 10^{48}~\mathrm{erg}$. The total radiated energy during our observation is therefore $7\times 10^{49}~\mathrm{erg}$. Because of the limited LC coverage of our observation, the actual radiated energy should be much larger than $7\times 10^{49}~\mathrm{erg}$.

The bolometric luminosity we presented is obtained by fitting BB functions to the SEDs from photometry.
When we have both spectra and photometry, we compare the results of the BB fits from spectra and photometry.
We found that the uncertainty from our use of the photometry to fit the BB function is up to 10\%. The systematic uncertainty caused by the use of the two BB functions to fit the SEDs is, however, uncertain.

\subsection{Spectroscopy}\label{sec:spectroscopy}

Plotted in Fig.~\ref{fig:allspectra}, we present our visual-wavelength and NIR spectra  of SN~2014ab extending from +54~d to +183~d, relative to the first \textit{WISE} detection in the rest frame. The spectra are dominated by narrow emission lines, justifying a SN~IIn classification. 

Figure~\ref{fig:hprofiles} shows the temporal evolution of prevalent hydrogen emission-line profiles. The hydrogen emission-line profiles  exhibit  little evolution  over the first 120 days of spectroscopic observations. The broad component matches the Lorentzian function with a full-width half maximum (FWHM) velocity of $3500~\kmps$. This Lorentzian line profile is formed by the multiple electron scattering in the optically thick CSM
(e.g., \citealt{chugai2001ecbroad}, but see also \citealt{huang2018ecbroad}). 
In superposing the broad component, we find a narrow emission component as shown in the top panel of Fig.~\ref{fig:hprofiles}. The peak of the narrow H$\alpha$ component in the X-shooter spectra match well and are used to determine the redshift of SN~2014ab ($z=0.02262\pm 0.00001$). 
The difference between the host galaxy redshift ($z=0.02352$) and
the SN redshift ($z=0.02262$)
corresponds to the line-of-sight velocity difference of 270~\kmps. The difference may originate from the host galaxy's rotation and is within the normal distribution of radial velocity of SN host galaxies \citep{galbany2018pisco}.
We can also see a tentative absorption component in the narrow H$\alpha$ lines with a minimum at $-80~\kmps$.
This feature could be produced  from  an unshocked wind with a characteristic velocity of $80~\kmps$. Such an absorption with the minimum velocity of arouund 100~\kmps\ is often observed in SNe~IIn \citep[e.g.,][]{taddia2013iin}.
The peak of the broad component is shifted by $-500~\kmps$. 

The broad emission lines show little  evolution during our spectroscopic observations (Fig.~\ref{fig:hprofiles}). The emission-line luminosities of the representative Balmer lines are $\simeq 7\times10^{41}~\mathrm{erg~s^{-1}}$ (H$\alpha$) and $\simeq 2\times10^{41}~\mathrm{erg~s^{-1}}$ (H$\beta$). The flux ratio of H$\alpha$ to H$\beta$ is $\simeq 4$, and is similar to that of SN~2010jl at around 100~days after the estimated explosion date \citep{fransson2014sn2010jl}. Similarly, the emission line luminosities of the representative Paschen lines are $\simeq 2\times10^{41}~\mathrm{erg~s^{-1}}$ (Pa$\alpha$) and $\simeq 9\times10^{40}~\mathrm{erg~s^{-1}}$ (Pa$\beta$). 
The luminosity of Pa$\alpha$ is difficult to measure because of the strong telluric absorption and we therefore measure it by using the +97~d spectrum which has a better telluric correction.

\section{Discussion}

\subsection{Comparison of SN~2014ab to other SNe~IIn}
We now compare SN~2014ab with other SNe~IIn. The explosion date of SN~2014ab is not well-constrained. Here, we arbitrarily assume that the explosion date is 100~days before the optical discovery in the rest frame which is 46~days before the first \textit{WISE} detection.
With this assumption on the explosion date, the LC of SN~2014ab matches that of SN~2010jl well, which is shown to be very similar to SN~2014ab later.
Figure~\ref{fig:lightcurvecomparison} presents the LC comparison. In the first 200~days, the LC of SN~2014ab declines slowly and it resembles slowly declining SNe~IIn such as SN~1988Z and SN~2010jl. SN~2014ab is significantly brighter than SN~1988Z and it is slightly fainter than SN~2010jl. The optical LC of SN~2014ab is found to be similar to that of KISS15s in the early phases \citep{kokubo2019kiss15s}. However, their MIR LCs have different properties (Fig.~\ref{fig:wise}).  Specifically, the MIR luminosity of KISS15s rises gradually for 700~days, while in the case of SN~2014ab, the MIR luminosity   increases suddenly at around $+$300~d and subsequently gradually declines.

One important characteristic that makes SN~2014ab distinct from other slowly declining SNe~IIn is the optical luminosity drop that occurred between +200~d and +400~d (Fig.~\ref{fig:lightcurvecomparison}). The optical LCs of slowly evolving SNe~IIn sometimes start to decline rapidly from several hundred days after the explosion. The optical LC drop displayed by SN~2014ab is the most abrupt among slowly declining SNe~IIn. The change in the optical LC decline rate in slowly-declining SNe~IIn is often accompanied by an IR luminosity increase. Indeed, the sudden optical luminosity decline in SN~2014ab follows after an increase in luminosity at MIR wavelengths.

Figure~\ref{fig:speccompandid} shows the spectral comparison of SN~2014ab with different SNe~IIn subtypes. This includes the slowly declining 1988Z-like Type~IIn SN~2010jl, the rapidly declining Type~IIn SN~1998S and, finally, the 1994W-like Type~IIn SN~2009kn with its LC  akin to normal SNe~IIP. The spectrum of SN~2010jl  closely resembles  that of  SN~2014ab, with both SNe having a very similar shape of their continua and the same spectral features.
Furthermore, the hydrogen-line flux ratios measured from the spectra of both objects are nearly identical. Comparison of the spectra of these two objects with that of  SN~2009kn reveals that the latter has much narrower emission-line features. Finally, the spectrum of SN~1998S is quite distinct from that of SN~2014ab. The spectral comparison also confirms that SN~2014ab is a member of the slowly declining, 1988Z-like subclass of SNe~IIn.

Inspection of the emission-lines associated with H$\alpha$ and the other hydrogen lines in SN~2014ab have the centroid of the broad emission line  blue-shifted  by 500~\kmps\ as shown in Fig.~\ref{fig:hprofiles}. A similar blue-shift of the order of 100~\kmps\ in the emission peak is also found in SN~2010jl, although the shift velocity is time-dependent \citep{gall2014sn2010jl,fransson2014sn2010jl}. In the case of SN~2014ab, we do not find the time dependent shift of the peak of the broad component as in SN~2010jl. In addition, we do not find the bluer emission peaks in bluer emission lines as observed in SN~2010jl. Indeed, the shape of the hydrogen emission lines in the spectra of SN~2014ab do not strongly depend on either emission wavelengths or time (Fig.~\ref{fig:hemissions}).

An interesting  difference found between the spectra of SN~2014ab and SN~2010jl is in the blended emission line of the  \ion{O}{i}~$\lambda8446$ and \ion{Ca}{ii} NIR triplet $\lambda\lambda8498, 8542, 8662$ features. These lines appear to be brighter in SN~2014ab than in SN~2010jl. The \ion{Ca}{ii} NIR triplet is likely produced within the cold dense shell where  the hydrogen emission lines  also originate \citep[e.g.,][]{dessart2016iin}. Since the strength of the hydrogen emission lines are similar in SN~2014ab and SN~2010jl, the difference in the emission strength found here is presumably due to the \ion{O}{i} line rather than the \ion{Ca}{ii} NIR triplet. Indeed, the emission peak is bluer in SN~2014ab. 
Since the emission from the inner SN ejecta is hidden by the electron scattering and no other \ion{O}{i} lines are observed, the \ion{O}{i}~$\lambda8446$ emission is likely to be excited by  Ly$\beta$ fluorescence \citep{bowen1947,bhatia1995,kastner1995}.

\subsection{Luminosity evolution}
As we have discussed so far, SN~2014ab is similar to SN~2010jl in many respects. Figure~\ref{fig:lightcurvemodels} shows a comparison between the bolometric LCs of SN~2014ab and SN~2010jl. The bolometric LC of SN~2014ab is obtained by integrating the two blackbody components shown in Section~\ref{bolometicLC} and the bolometric LC of SN~2010jl is obtained by integrating the photometry in optical ($3400-9000$~\AA) and NIR ($9000-24000$~\AA) as given in \citet{fransson2014sn2010jl}. We assumed that the explosion date of SN~2014ab is 100~days before the optical discovery, which provides a good match to the bolometric LC of SN~2010jl. The explosion date, however, is uncertain.
Under this assumption, the bolometric LC can be fit with the same power-law function as in SN~2010jl with a different scaling factor \citep{fransson2014sn2010jl}:
\begin{equation}
    L=1.2\times 10^{44}\left(\frac{t}{1~\mathrm{day}}\right)^{-0.54}~\mathrm{erg~s^{-1}},\label{eq:intlumfit}
\end{equation}
which is plotted in Fig.~\ref{fig:lightcurvemodels}.

Assuming that the mass loss from the progenitor is steady, the CSM density structure can be expressed as
\begin{equation}
    \rho_\mathrm{CSM}(r) = \frac{\dot{M}}{4\pi V_\mathrm{wind}}r^{-2},
\end{equation}
where $\dot{M}$ is the mass-loss rate and $V_\mathrm{wind}$ is the wind velocity.
The bolometric luminosity from the shock interaction with the shock velocity of $V_\mathrm{shock}$ can be expressed as
\begin{equation}
    L = \frac{\epsilon}{2}\frac{\dot{M}}{V_\mathrm{wind}}V_\mathrm{shock}^3,
\end{equation}
where $\epsilon$ is the conversion efficiency from kinetic energy to radiation energy \citep[e.g.,][]{moriya2013iin}. The mass-loss rate, therefore, can be estimated as
\begin{equation}
    \dot{M}=0.3 \left(\frac{\epsilon}{0.3}\right)^{-1}\left(\frac{L}{10^{43}~\mathrm{erg~s^{-1}}}\right)\left(\frac{V_\mathrm{shock}}{3000~\mathrm{km~s^{-1}}} \right)^{-3}\frac{V_\mathrm{wind}}{80~\mathrm{km~s^{-1}}}\ \mathrm{\Msun~yr^{-1}}.\label{eq:masslossrate}
\end{equation}
The bolometric luminosity of SN~2014ab is around $10^{43}~\mathrm{erg~s^{-1}}$ at around $100-200$~days (Fig.~\ref{fig:lightcurvemodels}). Although $V_\mathrm{shock}$ is not constrained in SN~2014ab, well-observed SNe~IIn including
SN~2010jl \citep{fransson2014sn2010jl} and
SN~2013L \citep{taddia2020} 
typically show $V_\mathrm{shock}\simeq 3000~\mathrm{km~s^{-1}}$. 
The conversion efficiency, $\epsilon$, is another uncertain parameter. Numerical simulations often indicate that this parameter  is on the order of 0.1  \citep{vanmarle2010,moriya2013sn06gy}. It has also been suggested to be time dependent \citep{tsuna2019,takei2019}. For simplicity, we assume that $\epsilon$ is constant and adopt the value $\epsilon\simeq 0.3$, which is on the order of 0.1. These assumptions are reasonable enough for our order-of-magnitude estimate.
Assuming the wind velocity estimated obtained from the narrow H$\alpha$ P-Cygni profile ($\simeq 80~\kmps$, Section~\ref{sec:spectroscopy}), Eq.~(\ref{eq:masslossrate}) then implies a mass-loss rate 
of SN~2014ab's progenitor of 0.3~\Msunpyr. This estimated mass-loss rate is similar to that found by \citet{fransson2014sn2010jl} for SN~2010jl ($\sim 0.1~\Msunpyr$). We note that the mass-loss rate estimated here should be regarded as a very rough estimate, given the uncertain shock velocity information and $\epsilon$. It is an order-of-magnitude estimate.

A significant difference between SN~2014ab and SN~2010jl appears at the  late phases beginning  around 400~days after explosion. The bolometric luminosity (hot + warm in Fig.~\ref{fig:lightcurvemodels}) of SN~2014ab decreases by a factor of  $\simeq 7$ 
during the gap in observations between 200 to 400 days, while over the same period,  the bolometric luminosity (optical + NIR in Fig.~\ref{fig:lightcurvemodels}) of SN~2010jl decreases by a factor of $\simeq 1.5$.
One possible reason for this discrepancy could be due to differences in the  structure of their CSM. Because the interaction luminosity is proportional to $\dot{M}$ (Eq.~\ref{eq:masslossrate}), the luminosity decrease by a factor of 0.15 corresponds to the $\dot{M}$ decrease by a factor of 0.15. Thus, the mass-loss rate of the progenitor of SN~2014ab may have been around $10^{-2}~\Msunpyr$ and increased to the order of 0.1~\Msunpyr\ shortly before the explosion, while the progenitor of SN~2010jl maintained the large mass-loss rate for a longer time. 
The difference in the mass-loss timescale may originate from the difference in the progenitor mass. The duration of extensive mass loss induced by pulsational pair-instability SNe, for example, strongly depends on the progenitor mass \citep[e.g.,][]{woosley2017ppisn}. 

We note that the late-phase bolometric luminosity decline is consistent with the nuclear decay rate of \Co\ (Fig.~\ref{fig:lightcurvemodels}). However, the required initial \Ni\ mass to account for the late-phase luminosity is 3~\Msun. If the ejecta have such a large amount of \Ni, it is likely to have an impact on the LC in the early phase because the total luminosity from the nuclear decay is larger than the observed luminosity over the entire early phase (Fig.~\ref{fig:lightcurvemodels}).
We also note that we cannot rule out the possibility that the majority of the radioactive energy  in the ejecta is thermalized and diffused with a long diffusion time to avoid the early effect on the luminosity. In such a case, however, the diffusion time in the ejecta needs to be more than 200~days following  ``Arnett's rule'' \citep{arnett1982law}, which would require an extremely large ejecta mass.
It may be argued that the NIR luminosity originates from light echoes and only the optical (hot) component should be regarded as the true bolometric luminosity. If this is the case, the initial \Ni\ mass required to account for the luminosity becomes 0.8~\Msun\ (Fig.~\ref{fig:lightcurvemodels}) and it does not affect the early luminosity much. However, this amount of \Ni\ is still much larger than would normally be synthesized in SNe \citep{anderson2019ni,meza2020ni}.

Finally, we  assumed that the dense CSM is spherically symmetric and comes from steady wind when analyzing the bolometric LC and deriving the mass-loss rate in this section. 
While we do not have clues on the CSM geometry in SN~2014ab, the dense CSM around SN~2010jl is likely to be aspherical \citep[e.g.,][]{patat2011sn2010jl}. We do not know whether or not if the mass loss is eruptive \citep[cf.][]{moriya2014iinhistory}. Despite of the possible asphericity and non-steady mass loss, \citet{fransson2014sn2010jl} estimate the mass-loss rate for SN~2010jl by assuming the spherical symmetry and steady mass loss for simplicity as we have done in this section. 
Thus, the estimated mass-loss rate of $\sim0.3~\Msunpyr$ should be, again, regarded as a rough estimate.

\subsection{Origin of dust}
Although SN~2014ab and SN~2010jl share many common characteristics, one big difference is in the signatures of dust. 
The IR excess indicates the existence of warm dust in both cases.
However, while the peak of the broad emission lines observed in SN~2010jl shifts towards shorter wavelengths over time \citep{smith2011sn10jl,gall2014sn2010jl,fransson2014sn2010jl}, this is not the case in SN~2014ab (Fig.~\ref{fig:hprofiles}). A shift of the broad emission peak to shorter (bluer) wavelengths is a signature of ongoing dust formation \citep[cf.][]{kozasa2009}. The peak of the broad hydrogen emission lines in SN~2010jl also shows the wavelength dependence expected from the dust absorption, that is, the peak of the emission lines with shorter wavelengths shifts more to the blue \citep{gall2014sn2010jl,maeda2013sn2010jl}. 
As demonstrated in Fig.~\ref{fig:hemissions}, this phenomenon is not observed in SN~2014ab.
The overall blue-shift of the broad emission lines can be explained by the acceleration of the unshocked CSM by precursor photons as suggested by \citet{fransson2014sn2010jl} for SN~2010jl (see also \citealt{moriya2011,chevalier2011irwin}).
It has also been suggested that blue-shift is affected by the radiation transfer effect in addition to the pre-shock acceleration \citep{dessart2015slsniin}.
SN~2014ab can be a good object to investigate the blue-shift caused by the hydrodynamic and radiation transfer effects because it may be less affected by dust as we discuss below.

The fact that we do not see clear observational signatures of ongoing dust formation in SN~2014ab may indicate that the dust found to be associated with it is not formed within the cool, dense shell where the emission lines are formed during our early-phase observations. The $BB_{warm}$ component does not exist at early phases. At late phases with $T^{BB}_{warm}$ estimated to be $\sim 2000~\mathrm{K}$, we find that this temperature is too high to originate from the inner SN ejecta located below the cold dense shell \citep[e.g.,][]{wooden1993,nozawa2003}. Indeed, $T^{BB}_{warm}\sim 2000~\mathrm{K}$ is close to the dust evaporation temperature.
Therefore, the pre-existing dust in the dense CSM may be heated by the SN shock or radiation, which drives an increase in the  MIR luminosity \citep{fox2010}. Interestingly, the MIR luminosity keeps decreasing after the sudden MIR luminosity increase that occurs between 200~days and 300~days (Fig.~\ref{fig:wise}). In the case of KISS15s, shown in the same figure, the MIR luminosity keeps increasing with the increasing mass of the presumably newly formed dust. Although we cannot rule out the possibility that dust is newly formed in the late phases, the MIR luminosity decline in SN~2014ab  is probably indicative  of little to no ongoing dust formation.  The decline in the MIR luminosity could be explained by the destruction of the pre-existing dust   by the SN shock wave as the ejecta expands over time.

In summary,  dust formation in SN~2014ab appears to be inefficient and the observed emission seems to be associated with pre-existing dust formed in the dense CSM prior to the final demise of the progenitor star.  SN~2014ab indicates that the origin of dust emission observed in SNe~IIn may not always be due to newly formed dust but rather sometimes could be due to  pre-existing dust. The exact reason why  dust formation in SN~2014ab is inefficient is not clear and warrants further  investigation.
Finally, we note that we cannot exclude the possibility that newly-formed dust may contribute to the late-phase MIR radiation. Indeed, both newly formed and pre-existing dust can appear at the same time, as has been suggested for SN~2006jd \citep{stritzinger2012iin} and other SNe~IIn \citep[e.g.,][]{gerardy2002,fox2009}.

\section{Conclusions}
We have reported the optical and NIR observations of Type~IIn SN 2014ab by CSP-II, as well as the serendipitous MIR observations by \textit{WISE}. Although the CSP-II observations started immediately after the optical discovery, the MIR observation showed that SN~2014ab had already been bright at 56~days before the optical discovery and the explosion date of SN~2014ab is uncertain. Nonetheless, we find that SN~2014ab shows  similarities with SN~2010jl, one of the best-observed slowly declining SNe~IIn. SN~2014ab is fainter than SN~2010jl by a factor of $\sim 2$, but its LC evolution is similar to that of SN~2010jl. The spectroscopic properties are almost identical to those of SN~2010jl. Based on the bolometric LC and assuming a spherically symmetric dense CSM and steady mass loss, the mass-loss rate of the progenitor immediately before the explosion is estimated to be on the order of 0.1~\Msunpyr.  

The MIR observation of SN~2014ab revealed that dust emission with the blackbody temperature of around 2000~K (close to the dust evaporation temperature) exists from around 350~days after the \textit{WISE} first detection. The MIR luminosity increases up to 350~days and then  declines afterwards. 
Although we observe the warm dust emission at late phases, no clear observational signatures of  dust formation within the cool dense shell such as the blue-shift of the broad line emission peaks with time and the wavelength-dependent shift of the emission peak are observed at early phases. These features were observed in SN~2010jl. In addition, KISS15s, which is another SN~2010jl-like SN~IIn with serendipitous MIR observations by \textit{WISE}, showed a gradual increase in the MIR luminosity, rather than the continuous decline found in SN~2014ab. Based on these differences, we presume that the dust emission observed in SN~2014ab originates from pre-existing dust located within the dense circumstellar matter heated by the SN shock or radiation, and the dust formation at the cool dense shell is much less  efficient in SN~2014ab as compared with SN~2010jl; however, the possibility of some contribution from newly-formed dust in the late-phase radiation is not excluded. The continuous MIR luminosity decline of SN~2014ab may also indicate that the heated dust is continuously destroyed by the shock. 
Our study of SN~2014ab  reveals  a diversity of plausible  origins for dust emission and dust formation in SNe~IIn.

\begin{acknowledgements}
We thank the anonymous referee for the constructive comments that improved the quality of the manuscript. 
We  thank Morgan Fraser and Takaya Nozawa for important discussions, and Ryan Foley, Howie Marion and Bob Kirshner for obtaining  FIRE spectra of SN~2010jl. 
T.J.M. is supported by the Grants-in-Aid for Scientific Research of the Japan Society for the Promotion of Science (JP17H02864, JP18K13585, JP20H00174).
M.S. and S.H. are supported in part by generous grants (project numbers 13261 and 28021) from VILLUM FONDEN.
M.S. and F.T. are also funded by a project grant (8021-00170B) from the Independent Research Fund Denmark.
N.B.S. acknowledges support from the NSF through grant AST-1613455, and through the Texas A\&M University Mitchell/Heep/Munnerlyn Chair in Observational Astronomy.
C.G. was supported by a VILLUM FONDEN Young Investigator grant (project number 25501).
This work was supported by a VILLUM FONDEN Investigator grant to J.H. (project number 16599).
L.G. was funded by the European Union's Horizon 2020 research and innovation programme under the Marie Sk\l{}odowska-Curie grant agreement No. 839090. This work has been partially supported by the Spanish grant PGC2018-095317-B-C21 within the European Funds for Regional Development (FEDER).
This work is supported by the Japan Society for the Promotion of Science Open Partnership Bilateral Joint Research Project between Japan and Chile (JPJSBP120209937).
The CSP-II has been funded by the USA's NSF under grants AST-0306969, AST-0607438, AST-1008343, AST-1613426, AST-1613455, and AST-1613472, and in part by a Sapere Aude Level 2 grant funded by the Danish Agency for Science and Technology and Innovation  (PI M.S.).
NTT (+ ELFOSC) spectrum was obtained by ESO program 191.D-0935.
VLT (+ X-shooter) spectra were obtained under ESO program 092.D-0645, and the data were retrieved from the ESO archive.
This work made use of the Open Supernova Catalog 
(\citealt{guillochon2017osc};~\url{https://sne.space/}) and WISeREP  (\citealt{yaron2012wiserep};~\url{https://wiserep.weizmann.ac.il}).
This research has made use of the NASA/IPAC Extragalactic Database (NED), which is operated by the Jet Propulsion Laboratory, California Institute of Technology, under contract with the National Aeronautics and Space Administration.
\end{acknowledgements}


\bibliographystyle{aa} 
\bibliography{references.bib} 

\begin{appendix}
\label{appendix1}

\section{Photometric data}
Photometric data of SN~2014ab used in our analysis are provided in this appendix. This includes two-channel WISE photometry of SN~2014ab listed in Table~\ref{wisephoto}, optical and NIR local sequences of SN~2014ab in Tables \ref{opticallocseq} and \ref{nirlocseq}, respectively, and optical/NIR photometry of SN~2014ab in Table~\ref{photometry}.

\clearpage
\input{tables/WISEphotometry}
\clearpage
\input{tables/opticallocsequence}
\input{tables/nirlocsequence}
\clearpage
\input{tables/photometry}
\end{appendix}

\begin{appendix}
\label{appendix2}
\section{CSP-II spectra of SN~2010jl}\label{sec:10jl}

\begin{figure*}
\centering
\includegraphics[width=1.8\columnwidth]{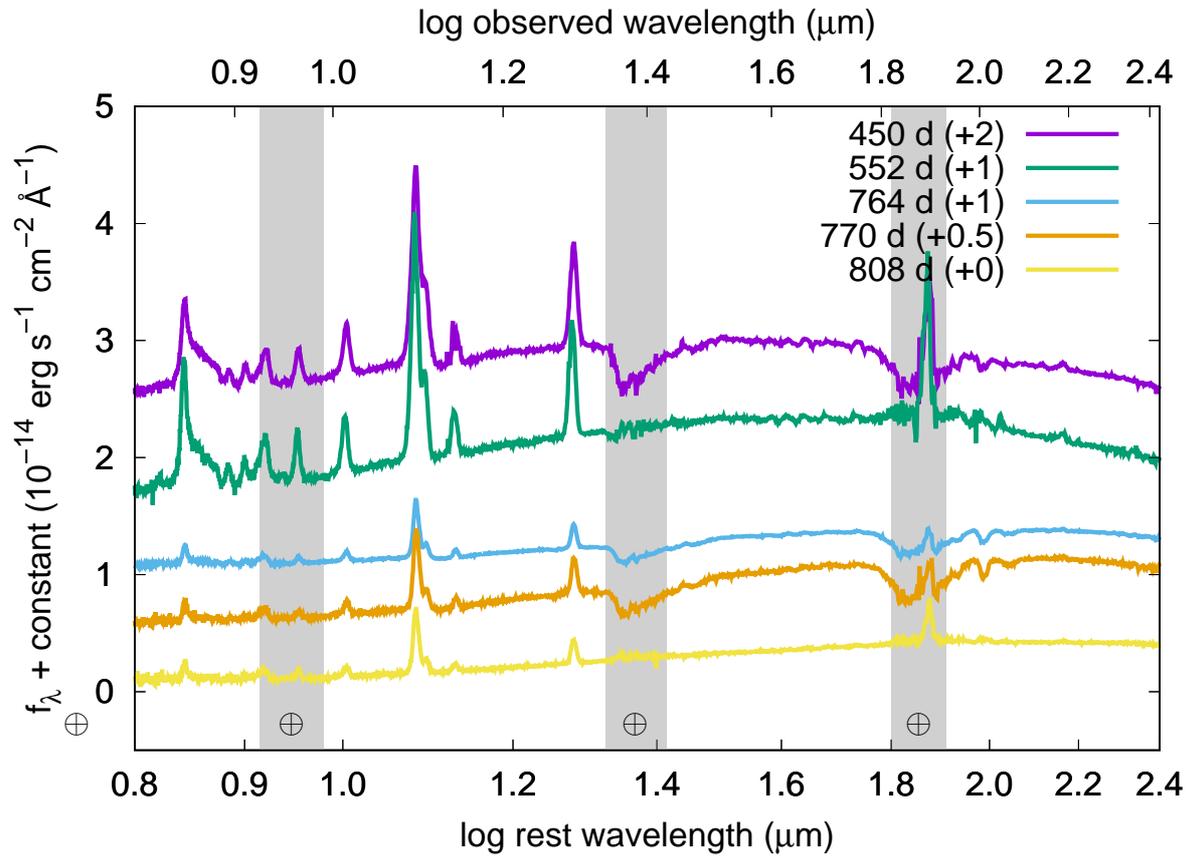} 
\caption{
NIR spectra of SN~2010jl obtained by the CSP-II. The epochs are in the rest frame relative to  the first detection which occurreed on JD = 2455479.0. The number in the parenthesis next to the epochs indicate the constant value added to each spectrum for presentation purposes and are   in the unit of $10^{-14}~\mathrm{erg~s^{-1}~cm^{-2}~\AA^{-1}}$. 
}
\label{fig:sn2010jlspec}%
\end{figure*}

In Fig.~\ref{fig:sn2010jlspec}, we present unpublished NIR spectra of SN~2010jl, which is shown in this paper to have common features with to SN~2014ab. A  journal summarizing the spectra obtained by the CSP-II of  SN~2010jl is provided in Table~\ref{specjorn10jl}.
The spectroscopic data was reduced following standard techniques. 
The visual-wavelength spectrum of SN~2010jl is shown in Fig.~\ref{fig:speccompandid}, while  five NIR spectra  obtained with the Baade (+ FIRE) telescope are plotted in Fig.~\ref{fig:sn2010jlspec}, and the rest-frame axes assumes the host-galaxy redshift of 0.0107.
Optical spectra of SN~2010jl at similar epochs have already been presented and analysed in the literature, but the NIR spectra cover late epochs than those published in the literature to our knowledge.
These data are presented here for the  benefit of the community. 

\clearpage

\clearpage
\input{tables/specjournal-2010jl.tex}

\end{appendix}

\end{document}

%% file: tables/specjournal.tex
\begin{table*}
\centering
\caption{Journal of spectroscopic observations of SN~2014ab.\label{specjor}}
\begin{tabular}{lccccc}
\hline\hline
Date (UT) & JD-2,456,000 & Phase$^a$&Telescope& Instrument & Range\\
&(days)&(days)&
&
&
(\AA) \\
\hline
10 Mar 2014 & 726.75 &  53.9   & Baade   & FIRE     & 8095--24810 \\
10 Mar 2014 & 726.81 &  54.0   & NTT     & EFOSC      & 3670--9260 \\
18 Mar 2014 & 734.79 &  61.8   & Baade   & FIRE     & 8095--24810\\
23 Mar 2014 & 739.73 &  66.6   & VLT     & X-shooter &3050--24790\\
25 Mar 2014 & 741.80 &  68.7   & Baade   & FIRE     & 8095--24810\\
23 Apr 2014 & 770.74 &  97.0   & Baade   & FIRE     & 8095--24810\\
24 Apr 2014 & 771.66 &  97.9   & VLT     & X-shooter &3050--24790\\
07 May 2014 & 784.72 &  110.6  & du Pont &   WFCCD    & 3630--9200\\
24 Jun 2014 & 832.54 &  157.4  & VLT     &   X-shooter   & 3050--24790 \\
29 Jun 2014 & 838.48 &  163.2  & VLT     &   X-shooter   & 3050--24790 \\
19 Jul 2014 & 858.48 &  182.7  & VLT     &   X-shooter   & 3050--24790 \\
\hline
\end{tabular}
\tablefoot{
$^{a}$Rest-frame days relative to the first \textit{WISE} detection.
}
\end{table*}

%% file: tables/WISEphotometry.tex
\begin{deluxetable}{cccc}
\tablewidth{0pt}
\tablecolumns{4}
\tablecaption{\textit{WISE} photometry of SN~2014ab in the Vega system.\label{wisephoto}}
\tablehead{
\colhead{Filter}&
\colhead{JD}&
\colhead{Magnitude}&
\colhead{Error}}
\startdata
\textit{W1} &   2456671.59 & 13.223 &  0.034 \\
&   2456671.72 & 13.163 &  0.031 \\
&   2456671.85 & 13.173 &  0.033 \\
&   2456671.98 & 13.094 &  0.038 \\
&   2456672.05 & 13.146 &  0.031 \\
&   2456672.11 & 13.194 &  0.031 \\
&   2456672.18 & 13.207 &  0.034 \\
&   2456672.18 & 13.187 &  0.046 \\
&   2456672.25 & 13.191 &  0.033 \\
&   2456672.31 & 13.099 &  0.044 \\
&   2456672.44 & 13.229 &  0.040 \\
&   2456672.57 & 13.216 &  0.035 \\
&   2456851.08 & 13.130 &  0.033 \\
&   2456851.21 & 13.108 &  0.039 \\
&   2456851.34 & 13.144 &  0.039 \\
&   2456851.47 & 13.126 &  0.039 \\
&   2456851.54 & 13.086 &  0.035 \\
&   2456851.60 & 13.124 &  0.031 \\
&   2456851.67 & 13.070 &  0.041 \\
&   2456851.74 & 13.134 &  0.031 \\
&   2456851.80 & 13.006 &  0.036 \\
&   2456851.93 & 13.124 &  0.032 \\
&   2456852.06 & 13.142 &  0.033 \\
&   2457035.48 & 12.526 &  0.024 \\
&   2457038.31 & 12.511 &  0.023 \\
&   2457038.44 & 12.478 &  0.022 \\
&   2457038.57 & 12.533 &  0.024 \\
&   2457038.63 & 12.397 &  0.024 \\
&   2457038.70 & 12.537 &  0.026 \\
&   2457038.77 & 12.472 &  0.023 \\
&   2457039.16 & 12.453 &  0.028 \\
&   2457039.16 & 12.503 &  0.028 \\
&   2457209.95 & 12.805 &  0.029 \\
&   2457210.09 & 12.807 &  0.029 \\
&   2457210.22 & 12.805 &  0.030 \\
&   2457210.28 & 12.809 &  0.030 \\
&   2457210.35 & 12.789 &  0.026 \\
&   2457210.41 & 12.811 &  0.028 \\
&   2457210.48 & 12.789 &  0.026 \\
&   2457210.54 & 12.837 &  0.030 \\
&   2457210.68 & 12.764 &  0.028 \\
&   2457210.81 & 12.739 &  0.026 \\
&   2457210.94 & 12.742 &  0.026 \\
&   2457400.64 & 13.174 &  0.032 \\
&   2457400.77 & 13.217 &  0.031 \\
&   2457400.91 & 13.219 &  0.039 \\
&   2457400.91 & 13.248 &  0.035 \\
&   2457401.10 & 13.129 &  0.032 \\
&   2457401.17 & 13.142 &  0.043 \\
&   2457401.23 & 13.185 &  0.034 \\
&   2457401.30 & 13.167 &  0.034 \\
&   2457401.36 & 13.221 &  0.036 \\
&   2457401.36 & 13.201 &  0.036 \\
&   2457401.50 & 13.323 &  0.040 \\
&   2457568.71 & 13.532 &  0.046 \\
&   2457568.84 & 13.529 &  0.046 \\
&   2457568.98 & 13.581 &  0.049 \\
&   2457569.11 & 13.656 &  0.046 \\
&   2457569.17 & 13.522 &  0.040 \\
&   2457569.24 & 13.440 &  0.038 \\
&   2457569.30 & 13.560 &  0.045 \\
&   2457569.37 & 13.470 &  0.039 \\
&   2457569.43 & 13.471 &  0.040 \\
&   2457569.70 & 13.616 &  0.040 \\
&   2457569.83 & 13.450 &  0.039 \\
&   2457764.80 & 13.738 &  0.047 \\
&   2457764.93 & 13.717 &  0.045 \\
&   2457765.06 & 13.896 &  0.072 \\
&   2457765.06 & 13.842 &  0.047 \\
&   2457765.19 & 13.848 &  0.071 \\
&   2457765.26 & 13.842 &  0.060 \\
&   2457765.39 & 14.008 &  0.069 \\
&   2457765.46 & 13.793 &  0.050 \\
&   2457765.52 & 13.760 &  0.050 \\
&   2457765.65 & 13.856 &  0.047 \\
&   2457765.78 & 13.937 &  0.054 \\
&   2457929.20 & 13.876 &  0.059 \\
&   2457929.33 & 14.027 &  0.053 \\
&   2457929.46 & 13.841 &  0.050 \\
&   2457929.53 & 14.095 &  0.052 \\
&   2457929.59 & 13.925 &  0.049 \\
&   2457929.66 & 13.933 &  0.070 \\
&   2457929.72 & 13.985 &  0.075 \\
&   2457929.92 & 13.919 &  0.047 \\
&   2457930.05 & 13.925 &  0.054 \\
&   2457930.18 & 14.048 &  0.056 \\
&   2457930.18 & 14.020 &  0.060 \\
&   2457936.99 & 13.957 &  0.057 \\
&   2458132.49 & 13.949 &  0.049 \\
&   2458132.56 & 14.049 &  0.053 \\
&   2458132.63 & 14.040 &  0.053 \\
&   2458132.69 & 13.920 &  0.064 \\
&   2458289.73 & 14.184 &  0.075 \\
&   2458289.99 & 14.193 &  0.056 \\
&   2458290.19 & 14.206 &  0.058 \\
&   2458290.58 & 14.152 &  0.067 \\
&   2458292.61 & 14.018 &  0.073 \\
&   2458293.07 & 14.033 &  0.055 \\
&   2458293.20 & 14.228 &  0.074 \\
\hline 
\textit{W2} &   2456671.59 & 12.716 &  0.055 \\
&   2456671.72 & 12.801 &  0.073 \\
&   2456671.85 & 12.685 &  0.055 \\
&   2456671.98 & 12.654 &  0.069 \\
&   2456672.05 & 12.689 &  0.057 \\
&   2456672.11 & 12.731 &  0.055 \\
&   2456672.18 & 12.800 &  0.057 \\
&   2456672.18 & 12.478 &  0.074 \\
&   2456672.25 & 12.771 &  0.059 \\
&   2456672.31 & 12.606 &  0.065 \\
&   2456672.44 & 12.655 &  0.058 \\
&   2456672.57 & 12.775 &  0.072 \\
&   2456851.08 & 12.689 &  0.053 \\
&   2456851.21 & 12.732 &  0.061 \\
&   2456851.34 & 12.622 &  0.056 \\
&   2456851.47 & 12.704 &  0.074 \\
&   2456851.54 & 12.712 &  0.054 \\
&   2456851.60 & 12.529 &  0.053 \\
&   2456851.67 & 12.672 &  0.065 \\
&   2456851.74 & 12.710 &  0.054 \\
&   2456851.80 & 12.761 &  0.082 \\
&   2456851.93 & 12.683 &  0.058 \\
&   2456852.06 & 12.690 &  0.056 \\
&   2457035.48 & 12.054 &  0.051 \\
&   2457038.31 & 12.009 &  0.034 \\
&   2457038.44 & 12.054 &  0.040 \\
&   2457038.57 & 12.016 &  0.036 \\
&   2457038.63 & 12.011 &  0.036 \\
&   2457038.70 & 12.019 &  0.037 \\
&   2457038.77 & 11.995 &  0.044 \\
&   2457039.16 & 12.032 &  0.037 \\
&   2457039.16 & 12.011 &  0.037 \\
&   2457209.95 & 12.200 &  0.039 \\
&   2457210.09 & 12.176 &  0.052 \\
&   2457210.22 & 12.172 &  0.041 \\
&   2457210.28 & 12.218 &  0.039 \\
&   2457210.35 & 12.151 &  0.043 \\
&   2457210.41 & 12.131 &  0.041 \\
&   2457210.48 & 12.154 &  0.039 \\
&   2457210.54 & 12.299 &  0.050 \\
&   2457210.68 & 12.240 &  0.043 \\
&   2457210.81 & 12.161 &  0.039 \\
&   2457210.94 & 12.212 &  0.039 \\
&   2457400.64 & 12.476 &  0.046 \\
&   2457400.77 & 12.445 &  0.044 \\
&   2457400.91 & 12.449 &  0.042 \\
&   2457400.91 & 12.468 &  0.046 \\
&   2457401.10 & 12.500 &  0.047 \\
&   2457401.17 & 12.486 &  0.051 \\
&   2457401.23 & 12.457 &  0.051 \\
&   2457401.30 & 12.489 &  0.061 \\
&   2457401.36 & 12.569 &  0.047 \\
&   2457401.36 & 12.484 &  0.046 \\
&   2457401.50 & 12.430 &  0.054 \\
&   2457568.71 & 12.818 &  0.062 \\
&   2457568.84 & 12.727 &  0.054 \\
&   2457568.98 & 12.757 &  0.053 \\
&   2457569.11 & 12.810 &  0.055 \\
&   2457569.17 & 12.690 &  0.066 \\
&   2457569.24 & 12.675 &  0.053 \\
&   2457569.30 & 12.723 &  0.062 \\
&   2457569.37 & 12.758 &  0.055 \\
&   2457569.43 & 12.679 &  0.053 \\
&   2457569.70 & 12.729 &  0.055 \\
&   2457569.83 & 12.764 &  0.056 \\
&   2457764.80 & 13.053 &  0.070 \\
&   2457764.93 & 13.050 &  0.077 \\
&   2457765.06 & 13.018 &  0.067 \\
&   2457765.06 & 13.088 &  0.083 \\
&   2457765.19 & 13.169 &  0.134 \\
&   2457765.26 & 12.990 &  0.067 \\
&   2457765.39 & 13.053 &  0.107 \\
&   2457765.46 & 12.942 &  0.077 \\
&   2457765.52 & 13.114 &  0.092 \\
&   2457765.65 & 13.032 &  0.074 \\
&   2457765.78 & 13.175 &  0.087 \\
&   2457929.20 & 13.260 &  0.101 \\
&   2457929.33 & 13.260 &  0.100 \\
&   2457929.46 & 13.350 &  0.108 \\
&   2457929.53 & 13.277 &  0.087 \\
&   2457929.59 & 13.132 &  0.105 \\
&   2457929.66 & 13.315 &  0.090 \\
&   2457929.72 & 13.111 &  0.114 \\
&   2457929.92 & 13.277 &  0.081 \\
&   2457930.05 & 13.314 &  0.114 \\
&   2457930.18 & 13.301 &  0.078 \\
&   2457930.18 & 13.485 &  0.094 \\
&   2457936.99 & 13.187 &  0.132 \\
&   2458132.49 & 13.337 &  0.136 \\
&   2458132.56 & 13.640 &  0.115 \\
&   2458132.63 & 13.546 &  0.114 \\
&   2458132.69 & 13.711 &  0.205 \\
&   2458289.73 & 13.480 &  0.117 \\
&   2458289.99 & 13.372 &  0.197 \\
&   2458290.19 & 13.551 &  0.229 \\
&   2458290.58 & 13.764 &  0.161 \\
&   2458292.61 & 13.612 &  0.226 \\
&   2458293.07 & 13.890 &  0.213 \\
&   2458293.20 & 13.265 &  0.140 \\
\enddata
\end{deluxetable}

%% file: tables/opticallocsequence.tex
\begin{deluxetable}{ccccccccc}
\tablewidth{0pt}
\tablecolumns{9}
   \tablecaption{Optical photometry of the local sequence for SN~2014ab in the `standard' system.\tablenotemark{a}\label{opticallocseq}}
   \tablehead{
      \colhead{ID} & 
      \colhead{$\alpha$ (2000)} & 
      \colhead{$\delta$ (2000)} & 
      \colhead{$B$} & 
      \colhead{$V$} & 
      \colhead{$u^{\prime}$} & 
      \colhead{$g^{\prime}$} & 
      \colhead{$r^{\prime}$} & 
      \colhead{$i^{\prime}$} }
      \startdata
      1 & 207.103256 &   7.430505 & 14.464(011)& 13.444(023)& 16.446(044)& 13.917(012)& 13.080(018)& 12.797(009)\\
  2 & 207.016541 &   7.351987 & 14.614(015)& 13.689(015)& 16.259(038)& 14.130(016)& 13.290(016)& 12.963(009)\\
  3 & 207.015869 &   7.294901 & 14.590(016)& 13.720(012)& 16.146(012)& 14.119(011)& 13.414(008)& 13.161(006)\\
  4 & 207.114792 &   7.291261 & 15.526(023)& 14.808(022)& 16.747(049)& 15.147(028)& 14.583(017)& 14.361(015)\\
  5 & 207.009033 &   7.314421 & 16.291(023)& 15.588(018)& 17.388(055)& 15.904(019)& 15.344(016)& 15.145(014)\\
  6 & 207.011932 &   7.279015 & 16.564(024)& 15.738(011)& 17.678(082)& 16.132(023)& 15.420(007)& 15.151(006)\\
  7 & 207.056549 &   7.408627 & 17.366(107)& 16.557(053)& 18.886(080)& 16.966(029)& 16.254(035)& 15.985(021)\\
  8 & 207.066711 &   7.456894 & 17.197(040)& 16.651(063)& 17.893(045)& 16.893(050)& 16.533(038)& 16.401(037)\\
  9 & 206.970734 &   7.467194 & 17.873(095)& 16.667(049)& 20.172(175)& 17.312(076)& 16.086(020)& 15.526(021)\\
 10 & 206.985626 &   7.469990 & 17.387(093)& 16.748(044)& 18.374(047)& 17.066(053)& 16.548(034)& 16.378(036)\\
 11 & 207.005112 &   7.462066 & 17.547(076)& 16.980(068)& 18.446(177)& 17.231(050)& 16.807(047)& 16.680(051)\\
 12 & 207.083344 &   7.439429 & 17.800(123)& 17.094(045)& 18.862(072)& 17.432(045)& 16.804(053)& 16.574(035)\\
 13 & 207.017670 &   7.366422 & 17.598(063)& 17.004(032)& 18.429(073)& 17.302(041)& 16.820(037)& 16.564(043)\\
 14 & 207.042816 &   7.346911 & 17.631(082)& 17.268(079)& 18.254(360)& 17.414(081)& 17.191(062)& 17.105(065)\\
 15 & 207.089340 &   7.359910 & 17.796(093)& 17.428(077)& 17.735(102)& 17.632(061)& 17.415(070)& 17.500(060)\\
 16 & 206.964035 &   7.465946 & 17.983(056)& 17.435(057)& 18.840(110)& 17.674(050)& 17.280(060)& 17.103(082)\\
 17 & 207.051956 &   7.360322 & 18.002(088)& 17.609(058)& 18.643(084)& 17.769(036)& 17.461(051)& 17.394(094)\\
 18 & 207.099777 &   7.315611 & 18.480(085)& 17.496(060)& \ldots     & 18.019(075)& 17.077(052)& 16.648(038)\\
 19 & 207.090088 &   7.375126 & 18.332(037)& 17.644(083)& 19.249(020)& 17.987(091)& 17.401(066)& 17.143(065)\\
 20 & 206.999039 &   7.336068 & 18.927(048)& 17.627(049)& \ldots     & 18.281(056)& 17.015(036)& 16.112(028)\\
 21 & 207.094254 &   7.321296 & 19.040(093)& 17.863(081)& \ldots     & 18.449(077)& 17.279(072)& 16.671(040)\\
 22 & 207.026749 &   7.289460 & 19.266(047)& 17.956(091)& \ldots     & 18.602(057)& 17.344(086)& 16.760(047)\\
 23 & 207.071671 &   7.423058 & 18.691(154)& 18.198(109)& 19.223(054)& 18.381(053)& 17.966(113)& 17.762(075)\\
\enddata  
\tablenotetext{a}{Note. -- Values in parenthesis are 1-$\sigma$ uncertainties that correspond to the rms of the instrumental errors of the photometry obtained over a minimum of three nights observed relative to standard star fields.}
\end{deluxetable}

%% file: tables/nirlocsequence.tex
\begin{deluxetable}{cccccc}
\tablecolumns{6}
\tablewidth{0pt}
\tablecaption{NIR photometry of the local sequences for SN~2014ab in the `standard' system.\tablenotemark{a}\label{nirlocseq}}
\tablehead{
\colhead{ID} &
\colhead{$\alpha (2000)$} &
\colhead{$\delta (2000)$} &
\colhead{$Y$} &
\colhead{$J$} &
\colhead{$H$} }
\startdata
101 & 13.803147 & 7.379118 & $14.173\pm0.011$ & $13.742\pm0.011$ & $13.132\pm0.004$\\ 
102 & 13.803267 & 7.379149 & $14.385\pm0.023$ & $14.036\pm0.006$ & $13.500\pm0.010$\\
103 & 13.802220 & 7.406726 & $15.159\pm0.007$ & $14.724\pm0.023$ & $14.238\pm0.005$\\
104 & 13.800127 & 7.382974 & $16.225\pm0.008$ & $15.775\pm0.029$ & $15.136\pm0.027$\\
105 & 13.802920 & 7.378393 & $16.685\pm0.067$ & $16.402\pm0.049$ & $15.866\pm0.058$\\
106 & 13.800973 & 7.380912 & $17.274\pm0.109$ & $16.859\pm0.085$ & $16.192\pm0.082$\\
\enddata
\tablenotetext{a}{Note. -- Values in parenthesis are 1-$\sigma$ uncertainties and correspond to an rms of the instrumental errors of the photometry obtained
over a minimum of three   photometric nights.}
\end{deluxetable}

%% file: tables/photometry.tex
\begin{deluxetable}{cccc}
\tablewidth{0pt}
\tablecolumns{4}
\tablecaption{Photometry of SN~2014ab in the `natural' system.\label{photometry}\tablenotemark{a}}
\tablehead{
\colhead{Filter}&
\colhead{JD}&
\colhead{Magnitude}&
\colhead{Error}}
\startdata
$u$ & 2456727.77 & 17.142 & 0.021      \\
    & 2456729.78 & 17.193 & 0.021      \\
    & 2456731.86 & 17.151 & 0.017      \\
    & 2456733.82 & 17.166 & 0.020      \\
    & 2456741.83 & 17.182 & 0.017      \\
    & 2456743.75 & 17.203 & 0.026      \\
    & 2456745.73 & 17.179 & 0.019      \\
    & 2456750.88 & 17.237 & 0.016      \\
    & 2456756.68 & 17.238 & 0.018      \\
    & 2456760.70 & 17.267 & 0.021      \\
    & 2456767.75 & 17.284 & 0.018      \\
    & 2456774.66 & 17.353 & 0.017      \\
    & 2456785.67 & 17.393 & 0.019      \\
    & 2456792.65 & 17.471 & 0.024      \\
    & 2456804.53 & 17.554 & 0.019      \\
    & 2456811.58 & 17.608 & 0.019      \\
    & 2457033.85 & 20.132 & 0.136      \\
\hline
$g$ & 2456727.76 & 16.439 & 0.017      \\
    & 2456729.78 & 16.437 & 0.016      \\
    & 2456731.86 & 16.447 & 0.012      \\
    & 2456733.81 & 16.469 & 0.016      \\
    & 2456741.82 & 16.494 & 0.010      \\
    & 2456743.74 & 16.496 & 0.013      \\
    & 2456745.72 & 16.486 & 0.012      \\
    & 2456750.87 & 16.538 & 0.011      \\
    & 2456756.67 & 16.528 & 0.015      \\
    & 2456760.69 & 16.553 & 0.016      \\
    & 2456767.75 & 16.589 & 0.017      \\
    & 2456774.66 & 16.619 & 0.011      \\
    & 2456785.67 & 16.684 & 0.017      \\
    & 2456792.65 & 16.732 & 0.013      \\
    & 2456804.52 & 16.802 & 0.012      \\
    & 2456811.58 & 16.844 & 0.016      \\
    & 2457031.84 & 20.180 & 0.162      \\
    & 2457033.84 & 20.136 & 0.056      \\
    & 2457065.87 & 20.715 & 0.089      \\
    & 2457095.82 & 20.824 & 0.092      \\
    & 2457126.69 & 21.331 & 0.091      \\
    & 2457156.73 & 21.963 & 0.198      \\
\hline
$r$ & 2456727.77 & 15.870 & 0.015      \\
    & 2456729.78 & 15.877 & 0.017      \\
    & 2456731.85 & 15.867 & 0.010      \\
    & 2456733.81 & 15.859 & 0.017      \\
    & 2456741.83 & 15.881 & 0.012      \\
    & 2456743.74 & 15.886 & 0.014      \\
    & 2456745.72 & 15.889 & 0.015      \\
    & 2456750.87 & 15.892 & 0.011      \\
    & 2456756.67 & 15.938 & 0.017      \\
    & 2456760.69 & 15.925 & 0.017      \\
    & 2456767.75 & 15.941 & 0.017      \\
    & 2456774.66 & 15.977 & 0.011      \\
    & 2456785.67 & 16.052 & 0.017      \\
    & 2456792.65 & 16.044 & 0.015      \\
    & 2456804.52 & 16.087 & 0.016      \\
    & 2456811.58 & 16.148 & 0.015      \\
    & 2457031.84 & 19.006 & 0.053      \\
    & 2457033.83 & 19.162 & 0.034      \\
    & 2457065.88 & 19.344 & 0.040      \\
    & 2457095.81 & 19.989 & 0.053      \\
    & 2457126.69 & 20.267 & 0.049      \\
    & 2457156.72 & 20.737 & 0.105      \\
\hline
$i$ & 2456727.77 & 16.114 & 0.021      \\
    & 2456729.78 & 16.135 & 0.024      \\
    & 2456731.85 & 16.142 & 0.011      \\
    & 2456733.82 & 16.134 & 0.020      \\
    & 2456741.83 & 16.167 & 0.014      \\
    & 2456743.74 & 16.221 & 0.026      \\
    & 2456745.72 & 16.181 & 0.015      \\
    & 2456750.87 & 16.223 & 0.010      \\
    & 2456756.67 & 16.262 & 0.022      \\
    & 2456760.69 & 16.294 & 0.023      \\
    & 2456767.75 & 16.317 & 0.021      \\
    & 2456774.66 & 16.357 & 0.013      \\
    & 2456785.67 & 16.431 & 0.020      \\
    & 2456792.65 & 16.538 & 0.021      \\
    & 2456804.52 & 16.551 & 0.019      \\
    & 2456811.58 & 16.614 & 0.018      \\
    & 2457033.84 & 19.691 & 0.065      \\
    & 2457065.88 & 19.790 & 0.081      \\
    & 2457156.73 & 20.805 & 0.180      \\
\hline
$B$ & 2456727.77 & 16.683 & 0.019      \\
    & 2456729.77 & 16.703 & 0.017      \\
    & 2456731.87 & 16.732 & 0.017      \\
    & 2456733.81 & 16.716 & 0.021      \\
    & 2456741.82 & 16.760 & 0.013      \\
    & 2456743.74 & 16.741 & 0.020      \\
    & 2456745.72 & 16.735 & 0.015      \\
    & 2456750.88 & 16.791 & 0.014      \\
    & 2456756.67 & 16.784 & 0.019      \\
    & 2456760.69 & 16.785 & 0.020      \\
    & 2456767.76 & 16.823 & 0.017      \\
    & 2456774.65 & 16.855 & 0.015      \\
    & 2456785.67 & 16.941 & 0.020      \\
    & 2456792.64 & 16.948 & 0.020      \\
    & 2456804.53 & 17.069 & 0.014      \\
    & 2456811.57 & 17.115 & 0.013      \\
    & 2457033.86 & 20.492 & 0.092      \\
    & 2457065.87 & 20.996 & 0.104      \\
    & 2457095.83 & 21.261 & 0.123      \\
    & 2457126.68 & 21.572 & 0.126      \\
\hline
$V$ & 2456727.77 & 16.319 & 0.021      \\
    & 2456729.77 & 16.319 & 0.016      \\
    & 2456731.87 & 16.328 & 0.013      \\
    & 2456733.81 & 16.315 & 0.017      \\
    & 2456741.82 & 16.341 & 0.012      \\
    & 2456743.74 & 16.340 & 0.015      \\
    & 2456745.72 & 16.369 & 0.014      \\
    & 2456750.88 & 16.390 & 0.013      \\
    & 2456756.67 & 16.387 & 0.016      \\
    & 2456760.69 & 16.420 & 0.018      \\
    & 2456767.76 & 16.454 & 0.016      \\
    & 2456774.65 & 16.475 & 0.012      \\
    & 2456785.67 & 16.553 & 0.019      \\
    & 2456792.64 & 16.619 & 0.017      \\
    & 2456804.53 & 16.673 & 0.015      \\
    & 2456811.57 & 16.723 & 0.014      \\
    & 2457031.84 & 20.000 & 0.153      \\
    & 2457033.85 & 20.148 & 0.081      \\
    & 2457065.87 & 20.643 & 0.107      \\
    & 2457095.82 & 20.763 & 0.104      \\
    & 2457126.69 & 21.059 & 0.122      \\
\hline
$Y$ & 2456727.81 & 15.290 & 0.008      \\
    & 2456731.82 & 15.302 & 0.007      \\
    & 2456762.74 & 15.395 & 0.007      \\
    & 2456786.67 & 15.566 & 0.011      \\
    & 2456796.65 & 15.586 & 0.007      \\
\hline
$J$ & 2456727.83 & 15.177 & 0.009   \\
    & 2456731.86 & 15.217 & 0.008   \\
    & 2456762.76 & 15.267 & 0.008   \\
    & 2456786.69 & 15.393 & 0.009   \\
    & 2456796.67 & 15.424 & 0.008   \\
\hline
$H$ & 2456727.83 & 15.002 & 0.008      \\
    & 2456731.85 & 15.055 & 0.007      \\
    & 2456762.75 & 15.126 & 0.008      \\
    & 2456786.69 & 15.226 & 0.009      \\
    & 2456796.66 & 15.273 & 0.009      \\
    \enddata
\tablenotetext{a}{Note. -- Uncertainties correspond to the 1-$\sigma$ instrumental magnitude uncertainty added in quadrature with the nightly zero-point uncertainty.}
\end{deluxetable}

%% file: tables/specjournal-2010jl.tex
\begin{deluxetable}{lccccl}
\tablewidth{0pt}
\tablecolumns{6}
\tablecaption{Journal of spectroscopic observations of SN~2010jl.\label{specjorn10jl}}
\tablehead{
\colhead{Date (UT)}&
\colhead{JD-2,455,000}&
\colhead{Phase\tablenotemark{a}}&
\colhead{Telescope}&
\colhead{Instrument}&
\colhead{Range}\\
\colhead{}&
\colhead{(days)}&
\colhead{(days)}&
\colhead{}&
\colhead{}&
\colhead{(\AA)}}
\startdata
04 Jan 2011 &  565.76 &  85.8  & du Pont & WFCCD    & 3630--9200\\
07 Jan 2012 &  933.75 &  449.9 & Baade   & FIRE     & 8095--24810 \\
19 Apr 2012 & 1036.54 &  551.6 & Baade   & FIRE     & 8095--24810 \\
19 Nov 2012 & 1250.83 &  763.7 & Baade   & FIRE     & 8095--24810 \\
25 Nov 2012 & 1256.82 &  769.6 & Baade   & FIRE     & 8095--24810 \\
03 Jan 2013 & 1295.82 &  808.2 & Baade   & FIRE     & 8095--24810 \\
\enddata
\tablenotetext{a}{Note. -- Rest-frame days relative to date of first detection, i.e., JD--2,455,479.0 \citep{fransson2014sn2010jl}.}
\end{deluxetable}